\documentclass[
superscriptaddress,
 amsmath,amssymb,
 aps,article,a4,
floatfix,floats,twocolumn
]{revtex4}

\usepackage{dcolumn}% Align table columns on decimal point
\usepackage{bm}% bold math
\usepackage[utf8]{inputenc}
\usepackage[T1]{fontenc}
\usepackage{amsmath}  
\usepackage{graphicx}
\usepackage{hyperref}
\usepackage{natbib}
\usepackage{ae}                         
\usepackage{verbatim}                   
\usepackage{subfigure}
\usepackage{pstool,graphicx}
\usepackage{epstopdf}
\usepackage{xcolor}

\newcommand{\beq}{\begin{equation}}
\newcommand{\eeq}{ \end{equation} }
\newcommand{\bea}{\begin{eqnarray}}
\newcommand{\eea}{\end{eqnarray}}
\newcommand{\mbf}{\mathbf}

\newcommand{\tk}{}

\usepackage{bm}% bold math

\newcommand{\Ma}{\rm{Ma}}
\newcommand{\Ra}{\rm{Ra}}
\newcommand{\Nu}{\rm{Nu}}
\newcommand{\Pe}{\rm{Pe}}
\newcommand{\Rey}{\rm{Re}}

\begin{document}

\title{Thermal Rayleigh-Marangoni convection in a three-layer liquid-metal-battery model}

\author{Thomas K\"ollner}
\altaffiliation[Present address: ]{Department of Mechanical Engineering,
University of California at Santa Barbara, Santa Barbara, California, USA}
\email{tk.koellner@googlemail.com}
\author{Thomas Boeck}
\email{thomas.boeck@tu-ilmenau.de}
\author{J\"org Schumacher}
\email{joerg.schumacher@tu-ilmenau.de}
\affiliation{Institut f\"ur Thermo- und Fluiddynamik, Technische Universit\"at Ilmenau, Postfach 100565, D-98684 Ilmenau, Germany}

\begin{abstract}
The combined effects of buoyancy-driven Rayleigh-B\'{e}nard convection (RC) and surface tension-driven Marangoni convection (MC) are studied in
a triple-layer configuration which serves as a simplified model for a liquid metal battery (LMB). The three-layer model consists of a liquid metal alloy 
cathode, a molten salt separation layer, and a liquid metal anode at the top. Convection is triggered by the temperature gradient between the hot 
electrolyte and the colder electrodes, which is a consequence of the release of resistive heat during operation.   We present  a linear stability analysis 
of the state of pure thermal conduction in combination with three-dimensional direct numerical simulations of the nonlinear turbulent evolution on the basis 
of a pseudospectral method. Five different modes of convection are identified in the configuration, which are partly coupled to each other: RC in the upper 
electrode, RC with internal heating in the molten salt layer, MC at both interfaces between molten salt and electrode as well as  anti-convection
in the middle layer and lower electrode. The linear stability analysis confirms that the additional Marangoni effect in the present setup increases the growth rates 
of the linearly unstable modes, i.e. Marangoni and Rayleigh-B\'{e}nard instability act together in the molten salt layer.
The critical Grashof and Marangoni numbers decrease with increasing middle layer thickness. The calculated thresholds for the onset of convection 
are found for realistic current densities of laboratory-sized LMBs. The global turbulent heat transfer follows scaling predictions for internally heated RC. 
The global turbulent momentum transfer is comparable with turbulent convection in the classical Rayleigh-B\'{e}nard case. In summary, our studies 
show that incorporating Marangoni effects generates smaller flow structures, alters the velocity magnitudes, and enhances the turbulent heat transfer 
across the triple-layer configuration. 
\end{abstract}

\date{\today}
\maketitle

\section{Introduction}
\label{sec:intro}

In the wake of the rapid growth of renewable energies, the intermediate storage of energy is of central importance. 
Besides thermal and mechanical methods, chemical energy storage in liquid metal batteries (LMB) is a promising 
way that received increasing attention in recent years \cite{kim2012liquid,wang2014lithium}. A liquid metal battery 
consists of three stratified  liquid layers. The heaviest layer is the cathode, which consists of a liquid metal alloy. It is 
separated from the liquid-metal  anode at the top by a  molten salt layer rather than by an ion-permeable solid separator 
as in standard batteries \cite{newman2012electrochemical}. This implies that the operating temperature of present 
prototypes is several hundreds degrees Celsius compared to room temperature in standard devices. The feasibility of 
such configurations was demonstrated recently in the laboratory. Among them are electrode combinations of lithium and 
lead-antimony alloys (Li||Pb-Sb) \citep{wang2014lithium} or magnesium and antimony (Mg||Sb) \citep{bradwell2012magnesium}, 
respectively.

LMBs are multi-physics fluid systems that couple thermal effects, magnetic fields and electrochemical reactions to turbulent 
flows in each of the three layers. Different aspects of LMB systems have been studied in the last years. The current-driven 
Tayler instability, which may cause an electrical short-circuit between both electrodes for large systems with diameters of the 
order of a meter, has been investigated by Weber et al. \cite{weber2013numerical,weber2015influence} and \citet{herreman2015tayler}. 
The metal pad instability, which is already known from aluminum reduction cells, has been studied by Zikanov \cite{zikanov2015metal}. 
Turbulent mixing processes in the liquid metal electrodes due to heating have been investigated experimentally by \citet{kelley2014mixing}. 
Very recently, thermal convection in the three-layer liquid metal battery was studied numerically by Shen and Zikanov \cite{shen2015thermal}. 
The authors showed in this work that in the presence of Joule heating, convective motion is always triggered for the typical Rayleigh or 
Grashof numbers in practical configurations. It can be expected that a significant part of the open physical questions and problems arises 
at the upper and lower interfaces between the electrodes and the molten-salt layer. This provides the motivation for our  study of 
interfacial convection in LMBs. Moreover, the temperature distribution is an important optimization parameter in the operation of LMBs 
since a high temperature leads to increased energy losses while a low temperature increases the likelihood of short circuits by solidification.

\tk{In the present work, we want to investigate the effects of interfacial tension in a LMB.   We ignore other processes in LMBs  
such as  mass transport and chemical reactions due to modeling uncertainties and in order to keep the  number of equations and parameters 
manageable.}
More precisely, we will  extend the three-layer thermal convection setting by Marangoni effects and study the linear stability and the full 
nonlinear evolution of such a LMB model. Our three-dimensional numerical simulations (DNS) will show that thermal Marangoni convection 
is an important factor for the flow inside a LMB. In this respect, we extend the model of \citet{shen2015thermal} to account for the 
differences of all transport coefficients and the gradient of interfacial tension. By doing this, interfaces are coupled by the continuity of all 
three velocities components across the interface and the balance of tangential viscous stresses with interfacial tension gradients. Effects 
of the electrical current and the resulting Ohmic heating are included by different heating rates in the three layers, which reflect the
differences in electrical conductivity. As a consequence, temperature gradients are produced that potentially cause interfacial tension 
driven flows (i.e. Marangoni convection) and buoyancy driven flows (i.e. Rayleigh-B\'{e}nard convection). Our numerical simulations will 
allow us to study these effects either in combination or independently, whereby their relative importance can be estimated.

The deformations of layers, which have been incorporated, e.g., in magnetohydrodynamic numerical simulations of \citet{herreman2015tayler}, 
are neglected. The azimuthal magnetic field, which is caused by the charge current across the cell is also neglected. This simplification  
is in line with the results of \citet{shen2015thermal}: \tk{The azimuthal magnetic field of a homogeneous current remains 
negligible for laboratory sized configurations which are in the focus of the present work, particularly in view to the nonlinear evolution of 
the three-layer model. The typical layer height is then of the order of centimeters.}  

Also, we will consider an internal fraction of the triple-layer only and exclude side walls. This will allow us to use the simpler Cartesian 
geometry in combination with periodic boundary conditions in the horizontal directions. The latter step opens the possibility to apply 
exponentially fast converging pseudospectral simulations \cite{koellner2016diss}. Finally, as in all previous studies, the model is 
considerably simplified by disregarding the mass transport of metal from the top layer to the bottom layer and the resulting change 
of density and interfacial tension with the composition. 

Thermophysical properties, especially for molten salts, are difficult to obtain from the literature for currently investigated LMBs. Thus we select the following representative substances for the LMB configuration: 

\begin{itemize}
\item For the upper electrode, we take lithium with a melting point at 181$^\circ$C. Lithium is a typical metal with favorable properties 
to construct LMBs \citep{kim2012liquid}. Material properties of lithium are taken from ref. \cite{maroni1973review}. 
\item For the layer that separates both liquid metal electrodes, a lithium chloride (59 mol\%)-potassium chloride (41 mol\%) eutectic 
mixture is taken which is denoted as LiCl-KCl.  The melting point is at 355$^\circ$C. Most of the required transport properties of eutectic 
LiCl-KCl are collected in \citet{williams2006assessment}. The surface tension value can be found in ref. \cite{smirnov1982density} and 
the electrical conductivity in \cite{van1955electrical}. The mass density as a function of temperature is given in ref. \cite{smirnov1982density}. 
It is noted that in \cite{shen2015thermal} the same salt used.
\item For the lower electrode, we take eutectic lead-bismuth (Pb-Bi), which has been also used in the mixing experiments by \citet{kelley2014mixing}. 
The material properties of eutectic Pb-Bi are also collected in ref. \cite{fazio2015handbook}. 
\end{itemize}
As a typical  temperature in an experiment  \citep{wang2014lithium} we take $T^{\infty}=500^\circ$C (773.15K), which is far above the melting 
temperature of all three battery components. The material properties at this operating temperature are collected in Tab.\ref{tab:physprop:ref}. 
The interfacial tension between layers is estimated by the method of \citet{girifalco1957theory}, see Sec.~\ref{sec:interfacial_tension}.

Our paper is organized as follows. Section II presents the non-dimensional model equations and the numerical method. 
Sec.~\ref{sec:stability} presents the linear stability analysis for the triple-layer configuration. Four convection regimes are identified: 
Rayleigh-B\'{e}nard convection (RC) in the middle and top layers, Marangoni convection (MC) at each of the two interfaces.  Based on linear 
stability investigations, Sec.~\ref{sec:nonlinear} discusses three-dimensional direct numerical simulation (DNS) of the full nonlinear evolution
at given  material parameters, which represent  typical laboratory-scale configurations in terms of currents and layer heights. In 
Sec.~\ref{sec:diss}, we derive a practical Rayleigh number criterion to decide when each of the four different regimes can appear and/or 
will dominate. \tk{The anti-convection mode (AC) in the bottom electrode is also considered.
Moreover, the relevance of interfacial deformations caused by the convection in the layers is estimated.} 
Finally, we summarize the present work in Sec.~\ref{sec:conclusion} and give a brief outlook.

\section{Three-layer model and numerical methods}

\subsection{Characteristic units}
\label{sec:governing:units}
\begin{figure}
\centering
\includegraphics[width=0.45\textwidth]{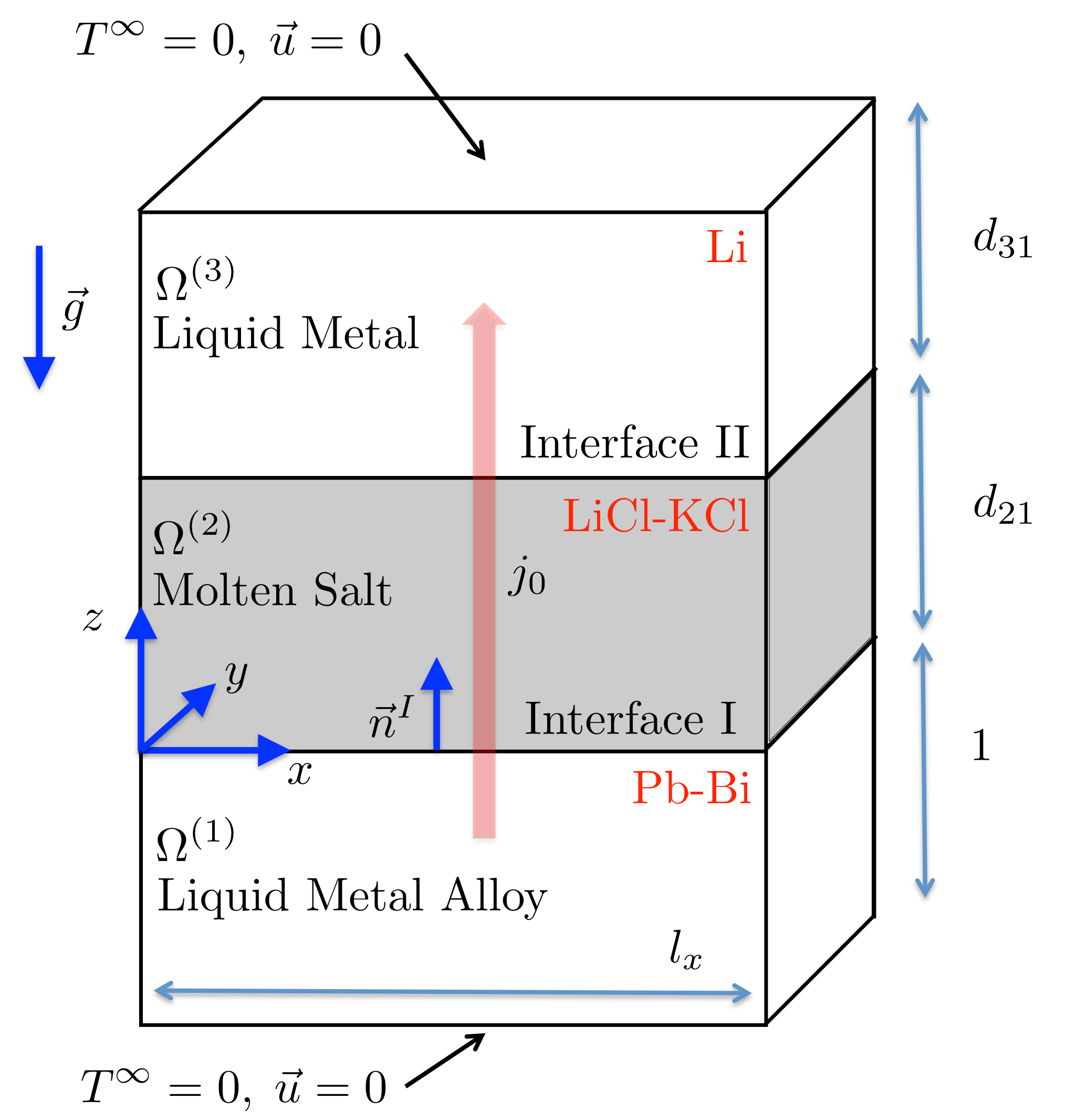}  
\caption{Sketch of the simplified three-layer liquid-metal battery model. All size lengths are given in units of the lower 
layer height $d^{(1)}$.}
\label{sketch}
\end{figure}

The model describes three immiscible Newtonian liquids that are stably stratified due to their density difference. 
Figure \ref{sketch} sketches this three-layer model in the non-dimensional formulation. 
The lower layer, representing a dense liquid-metal alloy, is located in the range $-d^{(1)}< z < 0$ and related 
quantities will be denoted with a superscript  $(1)$. The molten salt layer occupies the interval  $0< z<d^{(2)}$.  On top, a  
layer of liquid metal is located with $d^{(2)}<z<d^{(2)}+d^{(3)}$. Layer $\Omega^{(1)}$ and $\Omega^{(2)}$ are 
separated by the planar interface I, layer $\Omega^{(2)}$ and $\Omega^{(3)}$ by the  interface II, respectively.

The three layers are bounded by solid walls at the bottom and top, which are held at a uniform  temperature $T^{\infty}$. 
A homogenous internal heating is applied in each of the three layers, which mimics a prescribed homogeneous current 
density $j_0\mbf e_z$. The differences in the electrical conductivity $\sigma_e^{(i)}$ are also taken into account. 
Convection is driven by gradients in mass density $\rho^{(i)}(T)$ as well as the interfacial tensions $\sigma^{I}(T)$ 
and $\sigma^{II}(T)$, all of which depend on the temperature field $T(\mbf x, t)$.  

The characteristic units for time, $\tau_{vis}$, length, $L_{vis}$ and velocity, $U_{vis}$ are  based on the  height of the 
bottom layer and the characteristic time of viscous equilibration across this layer. They are given by
\beq
L_{vis}= d^{(1)}\,,\;\;\;
\tau_{vis}= \frac{\left(d^{(1)} \right)^2}{\nu^{(1)}}\,,\;\;\,
U_{vis}=\frac{\nu^{(1)} }{d^{(1)}}\,, 
\label{eq:theo:visscale:1}
\eeq
with $\nu^{(1)}$ being the kinematic viscosity in layer $\Omega^{(1)}$. The lateral aspect ratios in $x$-- and $y$--directions
are always equal, i.e. $l_x=l_y$. 
The appropriate temperature unit $\Theta$ is chosen to represent the maximum temperature appearing at pure conduction 
in the middle layer \citep{kulacki1972thermal}. It is given by
\beq
\Theta= \frac{Q^{(2)} (d^{(2)})^2  }{8\lambda^{(2)}}=\frac{ j_0^2 (d^{(2)})^2  }{8\lambda^{(2)} \sigma_e^{(2)}},
\label{Thetadef}
\eeq
where $Q^{(2)}= j_0^2/\sigma_e^{(2)}$ is the volumetric Joule dissipation rate (measured in J/(s m$^3$)) 
in the middle layer and $\lambda$ a thermal
conductivity. Also, $j_0$ is the constant current density and $\sigma_e$ is the electrical conductivity. Dimensionless
temperatures are thus given by $T=(\tilde T-T^{\infty})/\Theta$ where $\tilde T$ is the physical temperature. In Sec. 
\ref{sec:base_state}, the case of pure conduction is solved for the present three-layer problem. It will turn out that a 
zero heating rate in the top and bottom layer provides a good approximation to the exact conduction solution
since the electrical conductivity of the middle molten salt layer is considerably lower than in the outer liquid metal layers. 
Furthermore, the unit of temperature in eq. \eqref{Thetadef} will be a good estimate for the maximum temperature in the cell. 
Finally, the unit for pressure is given by  
\beq
P_{vis}=\rho_{ref}^{(1)}U_{vis}^2=\frac{\rho_{ref}^{(1)}\left(\nu^{(1)}\right)^2}{\left(d^{(1)}\right)^2}\,.
\eeq
From now on, we will consider dimensionless units only. The ratios of material parameters $\phi \in \{\nu, \kappa, \sigma, 
\lambda, \beta_T, ...\}$ are denoted by  the following abbreviation   
\beq
\phi_{ij}=\frac{\phi^{(i)}}{\phi^{(j)}}%\,.
\label{parameterratio}
\eeq 
They are summarized in Tab.~\ref{tab:nondim-parameters:ref}. This will simplify the notation of the set of equations 
which are discussed next.
%----------------------------------------------------
\begin{table}
\begin{center}
\begin{tabular}{lccc}
\hline\hline
Physical quantity & Symbol & SI-Unit & Value \\
\noalign{\smallskip}\hline\noalign{\smallskip}  
mass density & $\rho^{(1)}_{ref}$      		&  kg/m$^3$ 	& $1.0065\times 10^4$ \\
        & $\rho^{(2)}_{ref}$  		&  kg/m$^3$ 	& $1597.9$ \\
        & $\rho^{(3)}_{ref}$  		&  kg/m$^3$ 	& $484.7$ \\
kinematic viscosity &  $\nu^{(1)}$			& m$^{2}/$s 	&   $ 1.29  \times 10^{-7}$ 	\\ 
         &  $\nu^{(2)}$			& m$^{2}$/s 	& $1.38\times 10^{-6} $	\\
	 &  $\nu^{(3)}$			& m$^{2}$/s 	& $6.64\times 10^{-7} $	\\
thermal diffusivity 	&  $\kappa^{(1)}$		& m$^{2}$/s & $1.015\times 10^{-5}$	\\
	&  $\kappa^{(2)}$			& m$^{2}$/s & $1.90\times 10^{-7}$	\\
	&  $\kappa^{(3)}$			& m$^{2}$/s & $2.48\times 10^{-5}$	\\
thermal conductivity    	&  $\lambda^{(1)}$		& W/(mK)& 	14.41\\
	&  $\lambda^{(2)}$			& W/(mK) & 0.365	\\
	&  $\lambda^{(3)}$			& W/(mK) & 50.12	\\
specific heat 		&  $C_p^{(1)}$		& J/(kgK) & 141.05	\\
	&  $C_p^{(2)}$			& J/(kgK) & 1201.6	\\
	&  $C_p^{(3)}$			& J/(kgK) & 4169	\\
electrical conductivity	& $\sigma_e^{(1)}$  &  S/m&   $7.81\times 10^5$  \\
	& $\sigma_e^{(2)}$  &  S/m&  $187.1$  \\
	& $\sigma_e^{(3)}$  &  S/m&   $3 \times 10^6$ \\
expansion coefficient		&   $\beta^{(1)}_T$ &  1/K &  $-1.28\times 10^{-4}$   \\
					&   $\beta^{(2)}_T$ & 1/K & $ -3.32\times 10^{-4}$  \\
					&   $\beta^{(3)}_T$ & 1/K & $ -2.08\times 10^{-4}$ \\
change in interfacial 	&  $\sigma_{ref}^I \alpha_T^I$	& N/(mK)  &   $3.1\times 10^{-5}$	\\
tension	&  $\sigma_{ref}^{II} \alpha_T^{II}$	& N/(mK)  &   $7.98\times 10^{-5}$	\\
\noalign{\smallskip}\hline\hline\noalign{\smallskip}  
\end{tabular}
\caption{List of all material parameters of a three-layer battery system Li || LiCl-KCl || Pb-Bi
at an operating temperature of 500$^\circ$C. All references for the numerical values are given in the text.}
\label{tab:physprop:ref}
\end{center}
\end{table}
%----------------------------------------------------

\subsection{Equations and boundary conditions}
\label{sec:governing:eq}
In the following, we list the balance equations of momentum, mass and energy in nondimensional form. 
These equations have been successfully used to simulate Rayleigh-B\'{e}nard or Marangoni convection 
in other configurations \citep{nepomnyashchy2006interfacial,chilla2012new}. Since the magnitude of all material parameters
are related to those of the bottom layer, material parameter ratios appear additionally. The transport of momentum is 
given by the incompressible Navier-Stokes equations in the Boussinesq approximation 
\citep{nepomnyashchy2006interfacial,chilla2012new}
\begin{align}
\partial_t \mbf u^{(1)} =& -\left(\mbf u^{(1)} \cdot \nabla\right) \mbf u^{(1)} - \nabla p^{(1)}+ \nabla^2 \mbf u^{(1)} \nonumber \\
                                  &-G T^{(1)} \mbf e_z\,, \label{nse1} \\
\partial_t \mbf u^{(2)} = &-\left(\mbf u^{(2)} \cdot \nabla\right) \mbf u^{(2)} - \frac{1}{\rho_{21}} \nabla p^{(2)}+ \nu_{21} \nabla^2 \mbf u^{(2)} \nonumber\\
                                   &-G \beta_{21} T^{(2)} \mbf e_z\,, \\
\partial_t \mbf u^{(3)} =& -\left(\mbf u^{(3)} \cdot \nabla\right) \mbf u^{(3)} - \frac{1}{\rho_{31}} \nabla p^{(3)}+ \nu_{31} \nabla^2 \mbf u^{(3)} \nonumber\\
                                   &-G \beta_{31} T^{(3)} \mbf e_z\,,
\end{align}
with the Grashof number 
\beq
G=\frac{g\beta_T^{(1)}\left(d^{(1)}\right)^3\Theta}{\left(\nu^{(1)}\right)^2}=
\frac{g\beta_T^{(1)} j_0^2 \left(d^{(1)}\right)^3 \left(d^{(2)}\right)^2}{8\left(\nu^{(1)}\right)^2\lambda^{(2)} \sigma_e^{(2)}}\,.
\label{Grashof}
\eeq
Here, $\mbf g=(0,0,-g)$ is the vector of acceleration due to gravity and $\beta_T$ is the thermal expansion coefficient.
The mass balance is given by 
\begin{align}
\nabla \cdot\mbf u^{(i)}&=0\,,
\end{align}
with $i=1,2,3$. The variable $\mbf u$ denotes the velocity field. 
The energy balance reduces to a transport equation for the temperature field in each layer 
\begin{align}
\partial_t T^{(1)} =&- \left(\mathbf u^{(1)} \cdot \nabla\right) T^{(1)} \nonumber \\
                             &+ \frac{1}{\Pr^{(1)}} \Big [\nabla^2 T^{(1)} + \dfrac{8 \lambda_{21} \sigma_{e, 21}}{d_{21}^2}\Big ]\,,
\label{eq:thermal:lay1:nodim}\\ 
\partial_t T^{(2)} =&- \left(\mathbf u^{(2)} \cdot \nabla\right) T^{(2)} \nonumber \\
                            &+ \frac{\kappa_{21}}{\Pr^{(1)}} \Big [\nabla^2 T^{(2)} + \dfrac{8}{d_{21}^2}\Big ]\,,  
\label{eq:thermal:lay2:nodim} \\
\partial_t T^{(3)} =&- \left(\mathbf u^{(3)} \cdot \nabla\right) T^{(3)} \nonumber \\
                             &+ \frac{\kappa_{31}}{\Pr^{(1)}} \Big[\nabla^2 T^{(3)} + \dfrac{8 \lambda_{23} \sigma_{e, 23} }{d_{21}^2} \Big ]\,.  
\label{eq:thermal:lay3:nodim} 
\end{align}
The last term in each of the previous equations describes the non-dimensional volumetric heating rate.
The Prandtl number of the layer (i) is given by 
\beq
{\rm Pr}^{(i)}=\frac{\nu^{(i)}}{\kappa^{(i)}}\,.
\eeq
The Prandtl number of the middle layer is comparable to water whereas the metal layers have very small Prandtl numbers
of order $10^{-2}$, see Tab.~\ref{tab:nondim-parameters:ref}.
%--------------------------------
%--------------------------------

No-slip and  isothermal boundary conditions are imposed for the solid walls at the bottom and top:
\begin{align}
T^{(3)}&=0\,, \;\; \mbf u^{(3)}=0 \quad \text{for } z=d_{31}+d_{21}, \label{xx} \\
T^{(1)}&=0\,, \;\; \mbf u^{(1)}=0 \quad \text{for } z=-1. \label{xx}
\end{align}
The matching conditions at the lower planar interface I  at $z=0$ are as follows 
\begin{align}
\label{eq:matching:I:first}
u_x^{(1)}&= u_x^{(2)}\,, \ \ u_y^{(1)}= u_y^{(2)}\,, \\ 
u_z^{(1)}&= u_z^{(2)}=0\,, \\
\partial_{z} T^{(1)} &= \lambda_{21} \partial_{ z} T^{(2)}\,,\\ 
T^{(1)} &= T^{(2)}\,, \\
\frac{\Ma}{\Pr^{(1)}} \partial_{ x} T^{(1)} &= - \mu_{21} \partial_{ z} u_x^{(2)}+ \partial_{ z} u_x^{(1)}\,, \\
\frac{\Ma}{\Pr^{(1)}} \partial_{ y } T^{(1)} &= - \mu_{21} \partial_{ z} u_y^{(2)}+ \partial_{ z} u_y^{(1)}\,,
\label{eq:matching:I:last}
\end{align}
with the dynamic viscosities $\mu^{(i)}=\rho^{(i)}\nu^{(i)}$.
At the upper planar interface II at $z=d_{21}$, the matching conditions are as follows 
\begin{align}
\label{eq:matching:II:first}
u_x^{(2)}&= u_x^{(3)}\,, \ \ u_y^{(2)}= u_y^{(3)}\,, \\ 
u_z^{(2)}&= u_z^{(3)}=0\,, \\
\partial_{ z} T^{(2)} &= \lambda_{32}\partial_{ z} T^{(3)}\,, \\ 
T^{(2)} &= T^{(3)}\,, \\
\frac{\Ma\, \xi}{\Pr^{(1)}\tk{\mu_{21}}} \partial_{ x} T^{(2)} &= - \mu_{32} \partial_{ z} u_x^{(3)}+ \partial_{ z} u_x^{(2)}\,, \\
\frac{\Ma\, \xi}{\Pr^{(1)}\tk{\mu_{21}}} \partial_{ y } T^{(2)} &= - \mu_{32} \partial_{ z} u_y^{(3)}+ \partial_{ z} u_y^{(2)}\,.
\label{eq:matching:II:last}
\end{align}
These conditions \eqref{eq:matching:I:first}-\eqref{eq:matching:II:last} enforce the continuity of velocity components 
across the interfaces I and II,  the balance of the tangential stresses with interfacial tension gradients, and  a planar 
interface, which is an approximation to a more general normal-stress balance.  The variation of interfacial tension  
with temperature is given {\em in dimensional form} by
\begin{align}
\sigma^{I}&=\sigma_{\mbox{ref}}^{I}+\sigma_{\mbox{ref}}^{I} \,\alpha_T^{I} \left(\tilde T^{(1)}(z=0)- T^{\infty} \right)\,,
\label{eq:sigma:I}\\
\sigma^{II}&=\sigma_{\mbox{ref}}^{II}+\sigma_{\mbox{ref}}^{II} \,\alpha_T^{II} \left(\tilde T^{(2)}(z=d_{21}) -  T^{\infty} \right)\,,
\label{eq:sigma:II}
\end{align}
with $\alpha_T^{I}$ and $\alpha_T^{II}$ being the interfacial tension coefficients at both interfaces. The Marangoni number
is defined with respect to interface I and given by 
\beq
\Ma=\dfrac{\alpha_T^I j_0^2 \sigma_{ref}^{I} \left(d^{(2)}\right)^2 d^{(1)}}{ 8 \mu^{(1)} \kappa^{(1)} \lambda^{(2)}\sigma_e^{(2)}}
\eeq
The ratio of interfacial tension change with temperature between the  upper and the lower interface is quantified by 
\beq
\xi=\frac{\sigma_{\mbox{ref}}^{I}\,\alpha_T^{I}}{\sigma_{\mbox{ref}}^{II}\,\alpha_T^{II}}\,,
\label{interfaceratio}
\eeq
such that $Ma^{II}=\xi Ma$.
Table~\ref{tab:nondim-parameters:ref} summarizes all parameter values and ratios. The lateral boundary conditions are  
periodic, which is required by expanding the solution into a sum of  Fourier modes. Simulations are typically started with 
random velocity and zero temperature fields, representing the sudden switch-on of an electrical current.
%-----------------------------------------------------------
\begin{table}
\begin{tabular}{lll}
\hline\hline\noalign{\smallskip}
Quantity & Symbol & Value \\
\noalign{\smallskip}\hline\noalign{\smallskip}
Grashof number & $G$ & -3.97$\times 10^6$ \\
Marangoni number & Ma &  -310.02\\
Prandtl number      & Pr$^{(1)}$ & 0.0127 \\
                              & $Pr^{(2)}$ & 7.24 \\
                               & $Pr^{(3)}$ &  0.0268 \\
Layer height & $d_{21}$ & 1\\
                             & $d_{31}$ & 1\\
Thermal conductivity  & $\lambda_{21}$ & 0.0253\\
                            & $\lambda_{31}$ & 3.48\\
Mass density         & $\rho_{21}$ &0.159\\
                            & $\rho_{31}$ &0.048 \\
Kinematic viscosity & $\nu_{21}$ & 10.66 \\
                                        & $\nu_{31}$ & 5.14 \\
Thermal diffusivity & $\kappa_{21}$ & 0.019 \\
                                      & $\kappa_{31}$ & 2.44 \\
Thermal expansion & $\beta_{21}$ & 2.59\\
                                                & $\beta_{31}$ & 1.63 \\
Interfacial tension & $\xi$ & 2.57 \\
Electrical conductivity & $\sigma_{e, 21}$ & 2.396$\times 10^{-4}$ \\
                                           & $\sigma_{e, 31}$ & 3.841\\
Velocity unit	& $U_{vis}$ &  6.46$\times 10^{-6}$m/s \\
Time unit	& $\tau_{vis}$ &  3097 s \\
Length unit	& $L_{vis}$ &  20 mm \\
Temperature unit&  $\Theta$ & 6.59K\\
\noalign{\smallskip}\hline\hline
\end{tabular}
\caption{Derived non-dimensional parameters and parameter ratios which are used for the DNS of the nonlinear 
evolution. All ratios follow from (\ref{parameterratio}). Parameter $\xi$ is given by (\ref{interfaceratio}). 
Parameters that depend on layer heights and electrical current density are computed with
$d^{(1)}=d^{(2)}=d^{(3)}$=20 mm and $j_0=3$ kA/m$^{2}$.}
\label{tab:nondim-parameters:ref}
\end{table}
%-----------------------------------------------------------

\subsection{Estimation of interfacial tension}
\label{sec:interfacial_tension}
Most of the material properties required for our model are well documented. The appropriate references are given in 
Sec.~\ref{sec:intro}. However, the interfacial tension, a property that represents the molecular interaction between particles 
in the electrode and the molten salt \citep{rowlinson2002molecular}, is unknown for the particular case. Laboratory 
measurements of interfacial tension between liquid metals and molten salts are to the best of our knowledge not existent 
for the present configuration. 

To obtain interfacial tension, the method of \citet{girifalco1957theory} is employed here. It calculates interfacial tension as the sum 
of the well-known surface tensions of each component and subtracts the energy of the intermolecular bonds across the interface. 
Formally, it reads
\begin{align}
\sigma^{I}&= \sigma^{(1)}+\sigma^{(2)}-2\Phi^{I}\sqrt{\sigma^{(1)}\sigma^{(2)}}\,, \label{eq:girifalco1}\\
\sigma^{II}&= \sigma^{(2)}+\sigma^{(3)}-2\Phi^{II}\sqrt{\sigma^{(2)}\sigma^{(3)}}\,, \label{eq:girifalco2}
\end{align}
with the exchange parameter $\Phi \leq 1$. Values of $\Phi$  smaller than one account for molecular interactions that are different 
from dispersion type forces \citep{berg2010introduction}. Especially, for the present combination of ionic and metallic bonds, this 
parameter can be considerably different from unity. For the interface of aluminum and a salt mixture $\Phi=0.42$ has been noted 
by \citet{roy1998interfacial}). In the following, we take a value of $\Phi=0.7$.  

The following surface tension values $\sigma^{(i)}$ of each phase at an operating temperature of 500 $^\circ$C are extracted from  
the references in Sec.~\ref{sec:intro}. The particular values are $\sigma^{(1)}=0.387$ N/m [$\partial_T\sigma^{(1)}=7.99\times 10^{-5}N/(mK)$] 
for Pb-Bi, $\sigma^{(2)}=0.1327$ N/m [$\partial_T\sigma^{(2)}=8.26\times 10^{-5}N/(mK)$] for LiCl-KCl and $\sigma^{(3)}=0.3493$ N/m [$\partial_T
\sigma^{(3)}=1.6\times 10^{-4}N/(mK)$] for Li in the top electrode. With these values and Eqns.~(\ref{eq:girifalco1},\ref{eq:girifalco2})  
all free parameters in Eqns.~\eqref{eq:sigma:I} and \eqref{eq:sigma:II} can be calculated (see also Tab.~\ref{tab:physprop:ref}). Note 
also that the decrease of interfacial tension with temperature is in line with the fact that the mutual solubility of salt and liquid metal 
typically increases with temperature \cite{kim2012liquid}.

\subsection{Numerical method}
The numerical method is based on a pseudospectral algorithm using a Fourier expansion of all fields in $x$-- and $y$--directions 
with $N_x$ and $N_y$ modes, respectively. No-slip boundary conditions at the top and bottom as well as nonpenetrative boundary
conditions at the interfaces require a Chebyshev polynomial expansion with respect to $z$--direction with a polynomial degree of 
$N_z^{(i)}$ in each layer $\Omega^{(i)}$. For the polynomial degrees, we only use powers of two because of the fast Fourier 
transformations. The time-stepping scheme is a combination of the implicit Euler backward formula for linear terms and the Adams-Bashforth 
formula for nonlinear terms. The algorithm is a straightforward extension of the solver developed by \citet{boeck2002three},  
\citet{koellner2013multiscale} and \citet{koellner2016diss}.  The time step size is adapted such that the Courant-Friedrichs-Lewy number is 
in the range from 0.1 to 0.2.

The main difference to our former simulations \citep{koellner2013multiscale,koellner2016eruptive} is that now three instead of 
two layers are coupled. After discretization with respect to the horizontal $x$-- and $y$--directions as well as with respect to time, 
each individual Fourier mode of a hydrodynamic field satisfies a one-dimensional inhomogeneous Helmholtz equation  
\beq
\left(\frac{{\rm d}^2}{{\rm d}z^2} -\gamma^{(i)}\right) q^{(i)}(z)=f^{(i)}(z)
\eeq 
with $\gamma^{(i)}$ being a  real number, $f^{(i)}(z)$  a  given complex-valued given function, and $q^{(i)}(z)$  an unknown complex-valued 
function. These three  coupled equations (for $i\in\{1,2,3\}$) are solved directly with the Chebyshev tau method \cite{canuto1988spectral} 
in the same way as described  for two layers by \citet{boeck2002three}. The solver is written in the C language and parallelized using 
the Message Passing Interface library. The actual resolutions used are specified below.

The solver is validated by comparing with simple test cases, i.e., steady conduction and the Poiseuille flow that is driven by a homogeneous 
volume force in $x$-direction. Further checks were done by reproducing the stability threshold for the case treated by \citet{georis1999investigation}. 
We also reproduced the simulation results of \citet{shen2015thermal}, which however could be compared only qualitatively, because 
of different lateral boundary conditions. Nevertheless, flow structures were found to be rather similar. Furthermore, we checked that the kinetic 
energy balance holds and successfully reproduced the linear stability thresholds of convection, which is discussed in more detail in Sec.~\ref{sec:stability}.

\section {Linear stability analysis}
\label{sec:stability}
In the  following, we will investigate the linear stability of the stationary pure conduction state. 
The closed form of the stationary temperature distribution is detailed in the following Sec.~\ref{sec:base_state}. Thereafter, the linear stability 
problem is formulated (Sec.~\ref{sec:stab:problem}), its solution procedure is discussed, 
and critical Marangoni and Grashof numbers for the present reference system are calculated (Sec.\ref{sec:stab:results}). 

\subsection{Temperature profile of pure conduction}
\label{sec:base_state}

\begin{figure}
\centering	
\includegraphics[width=0.48\textwidth]{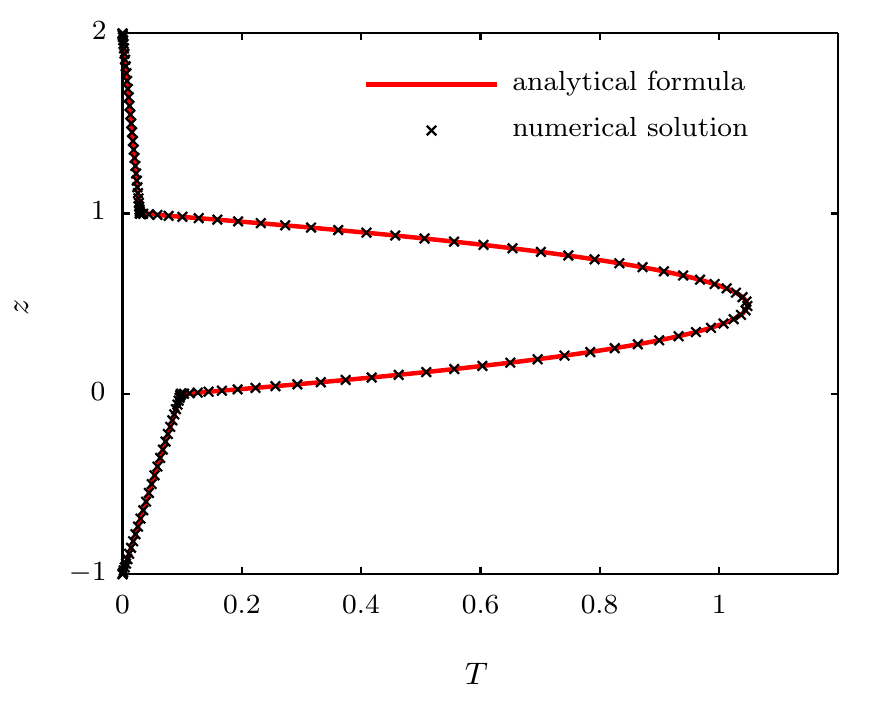} 
\caption{The stationary temperature profile of the reference case with the parameters taken from Tab.~\ref{tab:nondim-parameters:ref}.
In the pure conduction state one sets $Ma=G=0$. The profile is a plot of the solution \eqref{eq:conduction:internalheating:1}--
\eqref{eq:conduction:internalheating:3}. The profile is compared with the solution of a direct numerical simulation of the Boussinesq 
equations at the same parameters.}
\label{fig:conduction} 
\end{figure}

The case of pure conduction ($\mbf u=0$) with initial conditions $T^{(i)}(t=0)=0$ converges to a time-independent 
solution for $t\rightarrow \infty $ and can be found analytically. To further simplify the present problem, we neglect the 
Joule dissipation in the liquid metal electrodes and proceed with the assumption $\sigma_e^{(1)}=\sigma_e^{(3)}=\infty$, 
which will be justified later by comparison of the stability predictions with DNS of the fully nonlinear equations of motion. 
The steady state temperature distribution is a solution of Eqns.~\eqref{eq:thermal:lay1:nodim}--\eqref{eq:thermal:lay3:nodim}
and given by
\begin{align}
T^{(1)}_{cond}(z) &= \frac{4\lambda_{21} B}{d_{21} A}(1+z)\,, \label{eq:conduction:internalheating:1}\\
T^{(2)}_{cond}(z) &= \frac{4B}{A d_{21}}(\lambda_{21}+z)-\frac{4z^2}{d_{21}^2}\,, \label{eq:conduction:internalheating:2} \\
T^{(3)}_{cond}(z) &= \frac{4(d_{21}+2\lambda_{21})(d_{31}+d_{21}-z)}{A d_{21}} \label{eq:conduction:internalheating:3}\,,
\end{align}
where we introduced the abbreviations $A=d_{21}\lambda_{32}+\lambda_{31}+d_{31}$, $B=d_{21}\lambda_{32}+2d_{31}$.  
This temperature distribution is plotted together with the numerical solution that accounts for Joule dissipation in the liquid
metal layers in Fig.~\ref{fig:conduction}. One can readily observe that the top layer has higher conductivity than the lower one, 
resulting in a smaller temperature gradient in $\Omega^{(3)}$. Numerical and analytical solutions agree perfectly.

The temperature at both interfaces is given by 
\begin{align}
T^{I}_{cond}&=\frac{4B}{d_{21} A}\lambda_{21}\,, 
\label{eq:tempI}\\
T^{II}_{cond}&=T^I_{cond}\left(1+\frac{d_{21}}{\lambda_{21}}\right )- 4\,.  
\label{eq:tempII}
\end{align}
In the reference case, which is displayed in Fig.~\ref{fig:conduction}, the interfacial temperatures are  
$ T^{I}_{cond}=0.0995$ and $ T^{II}_{cond}=0.0296$. The maximum temperature appears in the middle layer. It is 
not exactly at $z=0.5$, but shifted slightly towards the layer with lower thermal conductivity. It occurs at
\beq
z^{max}=\frac{d_{21} B}{2A} \,, \label{eq:pos_max_temp}
\eeq
with a value of
\beq
T^{max}_{cond}=\frac{4B}{d_{21}A}\left(\lambda_{21}+z^{max}\right)-\left(\frac{B}{A}\right)^2\,. 
\label{eq:tempmax:cond}
\eeq
For the reference case, this results in $z_{max}=0.49$ and $T_{max}=1.06$. 
%-----------------------------------------------------------------
%-----------------------------------------------------------------

\subsection{Linearized equations and solution method}
\label{sec:stab:problem}

We proceed now with an extension of the linear stability analysis of Rayleigh-Marangoni convection to a three-layer model.
Former studies with a temperature difference applied between the boundaries  can be found in  
\cite{georis1993thermocapillary,nepomnyashchy2005convective,nepomnyashchy2006interfacial}. 
The vertical velocity component and the temperature perturbation are expanded into normal modes \cite{chandrasekhar1961hydrodynamic}
\begin{align}
 u_z^{(j)}(\mbf x,t)&= w^{(j)}(z,t)\exp(ik_x x+ ik_y y), \\ 
 T^{(j)}(\mbf x,t)&= \theta^{(j)}(z,t)\exp(ik_x x+ ik_y y)+T^{(j)}_{cond}(z).
\end{align}
Quantities $k_x, k_y$ are components of the horizontal wavenumber vector with magnitude $k=(k_x^2+k_y^2)^{1/2}$. 
The governing temperature equations are linearized around the basic state $T_{cond}(z)$. Furthermore,  by applying the 
curl two times to the linearized Navier-Stokes equations, the following system of linear equations \citep{nepomnyashchy2006interfacial} 
is derived 
\begin{align}
\partial_t[(D^2-k^2)w^{(1)}] &=    (D^2-k^2)^2 w^{(1)}+Gk^2\theta^{(1)}  \label{eq:dist:monotonic:w1},\\    	
\partial_t[(D^2-k^2)w^{(2)}]&=    \nu_{21}(D^2-k^{2})^{2} w^{(2)}\nonumber \\
                                          &+G\beta_{21} k^2\theta^{(2)} \label{eq:dist:monotonic:w2},\\
\partial_t[(D^2-k^2)w^{(3)}]&=    \nu_{31}(D^2-k^{2})^{2} w^{(3)}\nonumber \\
                                           &+G\beta_{31} k^2\theta^{(3)} \label{eq:dist:monotonic:w3}.
\end{align}
Here $D=d/dz$. For the temperature it follows that
\begin{align}
\partial_t \theta^{(1)}&=-w^{(1)} D T^{(1)}_{cond}  + \frac{1}{\Pr^{(1)}}(D^2-k^2) \theta^{(1)}, \label{eq:dist:monotonic:theta1}\\
\partial_t \theta^{(2)}&=-w^{(2)} D T^{(2)}_{cond}  + \frac{\kappa_{21}}{\Pr^{(1)}}(D^2-k^{2}) \theta^{(2)}, \label{eq:dist:monotonic:theta2}\\ 
\partial_t \theta^{(3)}&=-w^{(3)} D T^{(3)}_{cond}  + \frac{\kappa_{31}}{\Pr^{(1)}}(D^2-k^{2}) \theta^{(3)}. \label{eq:dist:monotonic:theta3}
\end{align}
The matching and boundary conditions at $z=0$ are given by 
\begin{align}
\theta^{(1)}&=\theta^{(2)},\\ 
D\theta^{(1)}&=\lambda_{21} D\theta^{(2)},  \label{eq:dist:match:1}\\ 
\mu_{21} D^2 w^{(2)} - D^2 w^{(1)} &= - \frac{\Ma^{I}}{\Pr^{(1)}} k^2 \theta^{(1)}, \label{eq:dist:match:2} \\
w^{(1)}&=w^{(2)}=0, \\  
D w^{(1)}&= D w^{(2)} \label{eq:dist:match:3}
\end{align}
and at $z=d_{21}$ by 
\begin{align}
\theta^{(2)}&=\theta^{(3)},\\
D\theta^{(2)}&=\lambda_{32} D\theta^{(3)},  \label{eq:dist:match:4}\\ 
\mu_{32} D^{2} w^{(3)}- D^2 w^{(2)} &= - \frac{\Ma^{II}}{\Pr^{(1)}\tk{\mu_{21}}} k^2 \theta^{(2)}, \label{eq:dist:match:5} \\
w^{(2)}&=w^{(3)}=0, \\  
D w^{(2)}&= D w^{(3)}, \label{eq:dist:match:6}
\end{align}
The boundary conditions at the no-slip walls are as follows. For $z=-1$
\beq
\theta^{(1)}=w^{(1)}=Dw^{(1)}=0,  \label{eq:dist:bot:1}\\ 
\eeq
and at $z=d_{21}+d_{31}$
\beq
\theta^{(3)}=w^{(3)}=Dw^{(3)}=0.  \label{eq:dist:top:1}\\ 
\eeq
Furthermore, we set $\Ma^I=\Ma$ and $\Ma^{II}=\xi \Ma$
in Eqns.~\eqref{eq:dist:match:2} and \eqref{eq:dist:match:5}, in order to study the  effect 
of Marangoni convection on both interfaces separately.

The resulting linear stability problem takes the form
\beq
\bm M\partial_t \mbf q(t,z) = \bm L \mbf q(t,z) \label{eq:linear:problem}
\eeq
where the column vector $\mbf q$ consists of the independent fields $\{ w^{(3)}, w^{(2)}, w^{(1)}, \theta^{(3)}, \theta^{(2)}, \theta^{(1)} \}$ 
and the linear differential operators $\bm M$ and $\bm L$ encode the bulk equations and the boundary conditions.  A general solution can be 
expressed by the exponential of the operator $(\bm M^{-1}\bm L)$ \citep{trefethen2005spectra}, which can be further analyzed to derive  
several properties of the solution. Also, one can expand the $\mbf q(z,t)$ into the eigenfunctions $\hat{ \mbf q_i}(z)$ of $(\bm M^{-1}\bm L)$ by 
$\mbf q= \sum_{i} \hat{ \mbf q_i}(z) \exp(t s_i)$ with  complex growth rates $s_i$.  We are interested  in the marginal stability properties only 
where $s_i=0$.  Then, the system can be rewritten as an eigenvalue problem with one control parameter. 

For the analysis of the onset of Marangoni convection, i.e. marginal stability, eq.~\eqref{eq:linear:problem}  with $\partial_t =0$ 
is discretized with the Chebyshev collocation method \cite{schmid2012stability,trefethen2000spectral} and rearranged thereafter 
to an eigenvalue problem with respect to the Marangoni number Ma. The result is a generalized eigenvalue problem with 
quadratic matrices $\bm A$ and $\bm B$ of dimension $6(N+1)\times 6(N+1)$, which follows to
\beq
{\Ma} \bm{A} q = \bm{B} q.  \label{eq:eigenvalue:problem:1}
\eeq
Here, $N$ is the degree of the Chebyshev polynomials which are used to discretize the vertical direction in each layer. It is usually set to 
$N=32$.  

%--------------------------------------------------
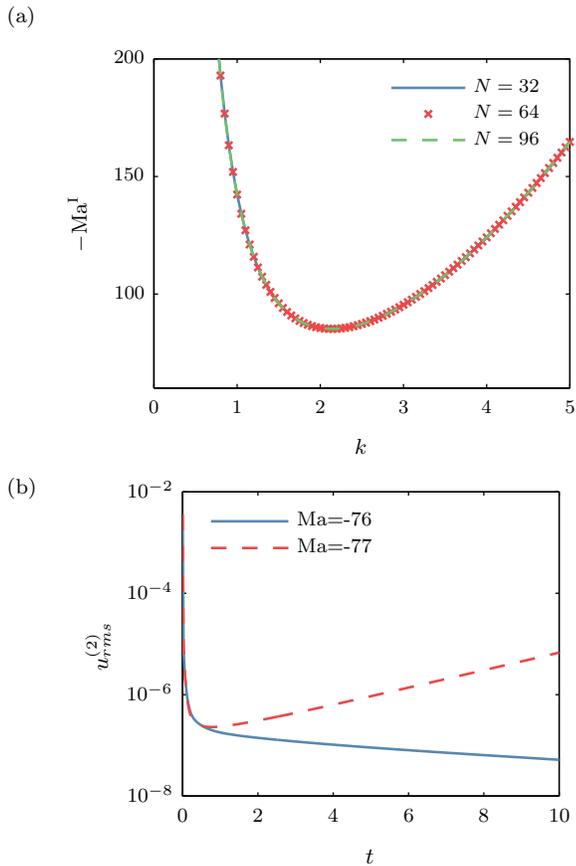
\begin{figure}
\setlength{\unitlength}{0.5\textwidth}
\includegraphics[width=0.4\textwidth]{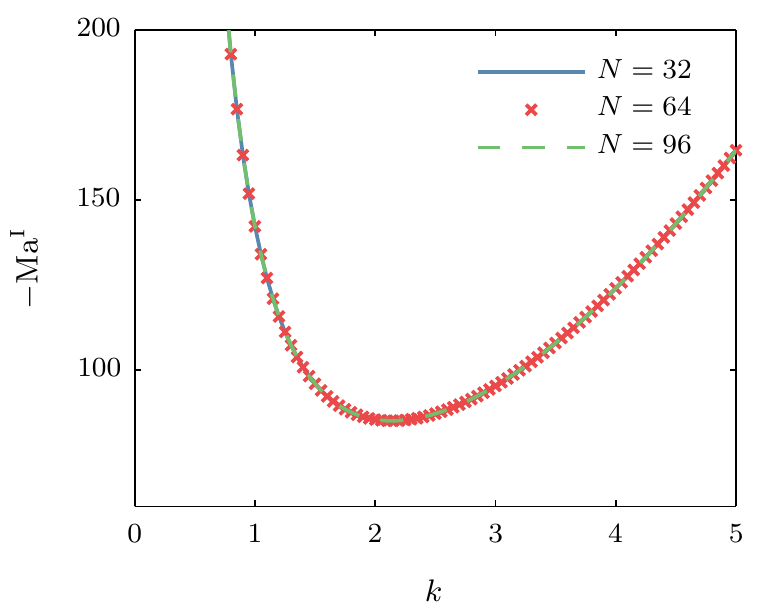} 
\includegraphics[ width=0.38\textwidth]{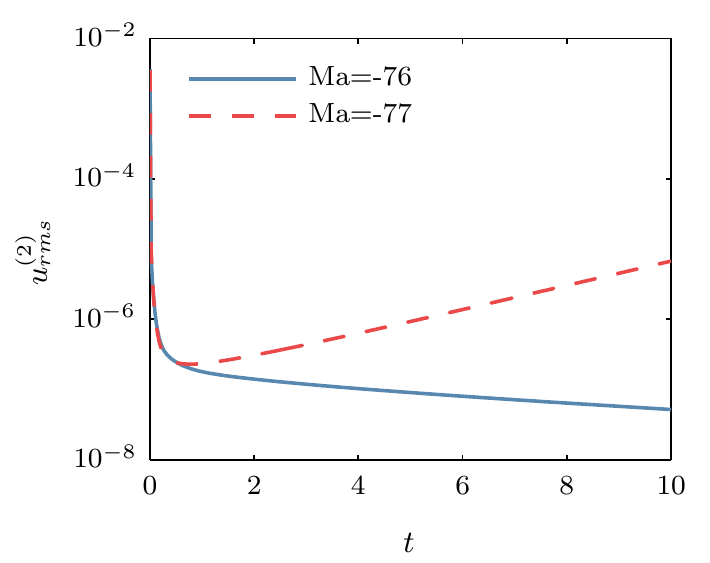} \\
\setlength{\unitlength}{0.5\textwidth}
\begin{picture}( 0, 0)(0,0)
\put(-0.48 , 1.3){\footnotesize (a)}
\put(-0.48 , 0.6){\footnotesize (b)}
\end{picture}
\caption{Linear stability analysis of the pure Marangoni convection (MC) case at $G=0$ for the reference parameters of 
Tab.~\ref{tab:nondim-parameters:ref}. Top (a): neutral stability curve $\Ma(k)$. Bottom (b): root mean square velocity 
in the mid layer for simulation runs LS01 and LS02.}
\label{fig:ma:stab:1} 
\end{figure}
%--------------------------------------------------

For a given set of parameters, the critical Marangoni number $\Ma_c$ is the eigenvalue to Eq.~\eqref{eq:eigenvalue:problem:1} with the 
minimal magnitude over all wavenumbers $k$ of the perturbations. Note that we only consider situations for which density and interfacial 
tension decrease with increasing temperature. This is equivalent to $\Ma\le 0$ and $G\leq 0$. Eigenvalues are calculated with {\it Matlab}. 
%--------------------------------------------------
\begin{figure}
\includegraphics[width=0.45\textwidth]{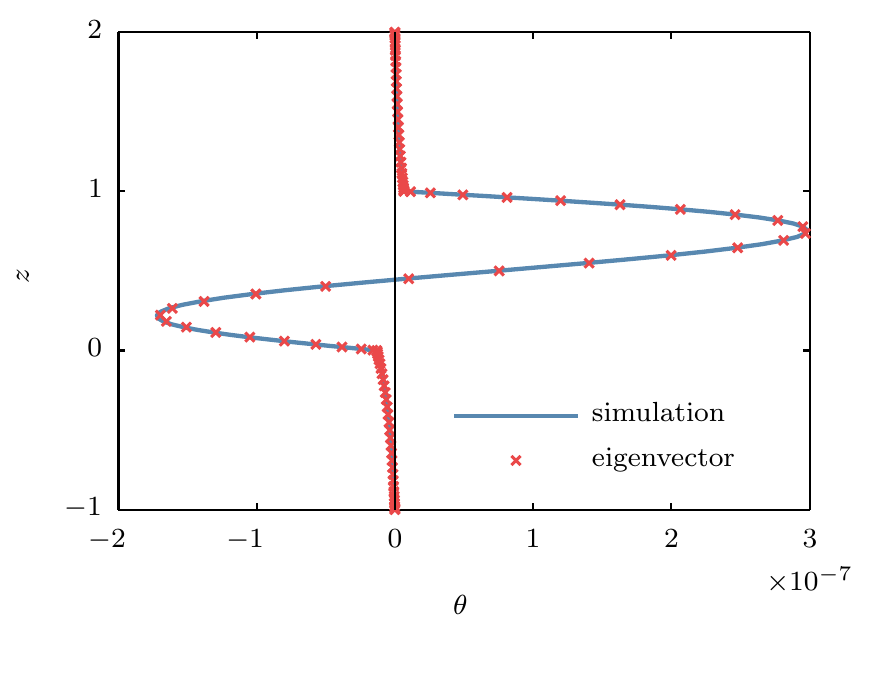}
\includegraphics[width=0.45\textwidth]{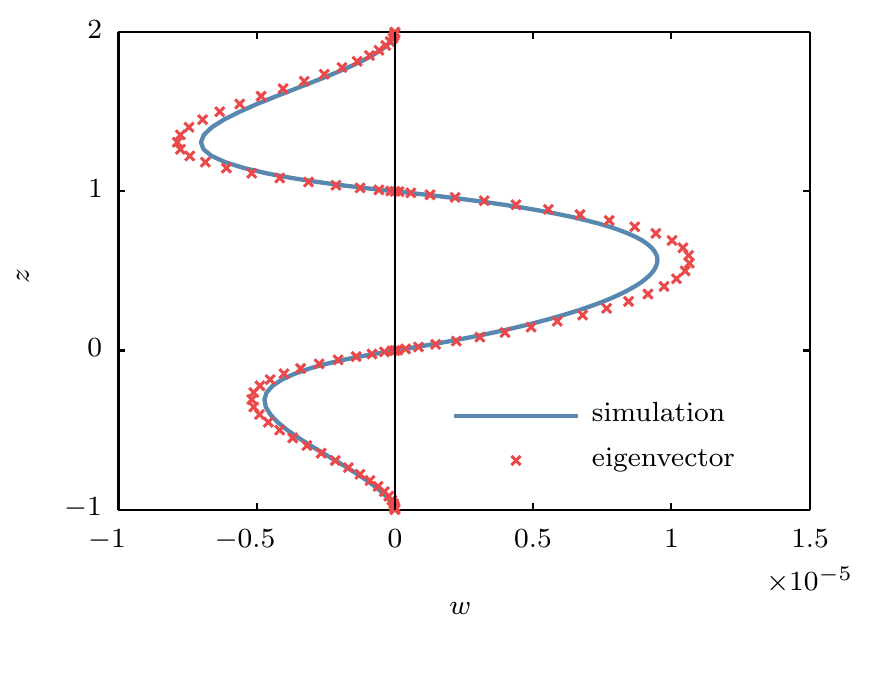}\\
\setlength{\unitlength}{0.5\textwidth}
\begin{picture}( 0, 0)(0,0)
\put(-0.48 , 1.40){\footnotesize (a)}
\put(-0.48 , 0.69){\footnotesize (b)}
\end{picture}
\caption{Comparison of eigenmodes from stability analysis and full numerical simulation. Top (a): the difference of 
the profiles of temperature, $T(x=0,y=0,z,t=10)-T(x=0,y=0,z,t=3)$, is plotted. Values from the eigenvalue calculation 
with  $k= 2.15$ and the simulation LS02 are shown. Bottom (b): the difference of the profiles of vertical velocity component, 
$u_z(x=0,y=0,z,t=10)-u_z(x=0,y=0,z,t=3)$ is displayed. Again, values from the eigenvalue calculation with  $k= 2.15$ 
and the simulation LS02 are shown.}
\label{fig:ma:stab:1a} 
\end{figure}
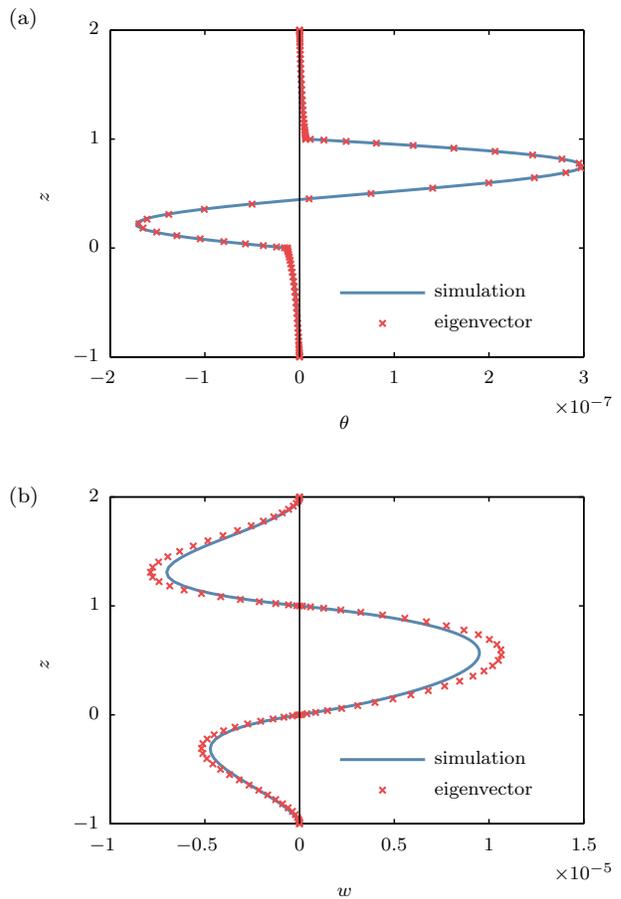
%--------------------------------------------------

\subsection{Results of  linear stability analysis}
\label{sec:stab:results}

\subsubsection{Pure Marangoni instability ($\Ma\ne 0$, $G=0$)} 
The neutral stability curve for the reference case with the parameters taken from Tab.~\ref{tab:nondim-parameters:ref} 
is displayed in Fig.~\ref{fig:ma:stab:1}(a). We find a critical value of \tk{$\Ma^I_c=-76.39$} at a wavenumber of \tk{$k_c=2.10$}. 
The marginal values of $\Ma^I$ (note that the coupling $\Ma^{II}=\xi \Ma^I$ applies here implicitly) were calculated for different 
polynomial degrees $N$. The critical value does not change up to the fourth digit.  
%-------------------------------------------------------
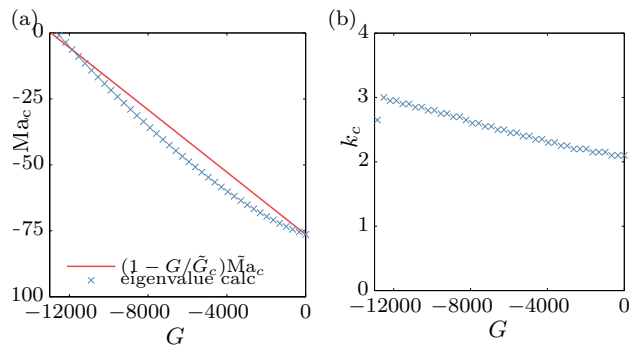
\begin{figure}
\includegraphics[width=0.23\textwidth]{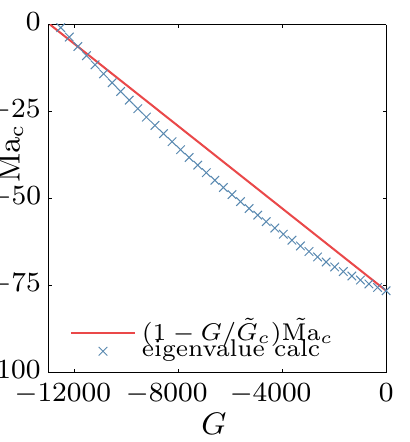}  
 \includegraphics[ width=0.23\textwidth]{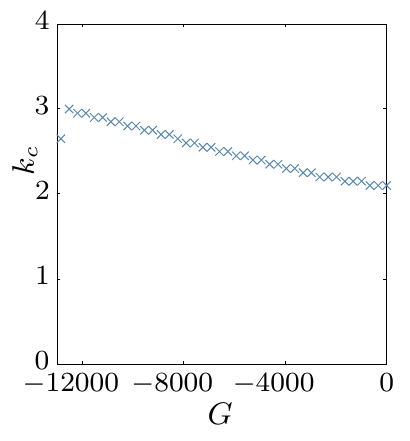}
\setlength{\unitlength}{0.5\textwidth}
\begin{picture}( 0, 0)(0,0)
\put(-0.96 , 0.48){\footnotesize (a)}
\put(-0.49 , 0.48){\footnotesize (b)}
\end{picture}
\caption{Linear stability analysis of Rayleigh-Marangoni convection (RMC) case at $\Ma\ne 0$ and $G\ne 0$. 
Left (a): critical Marangoni number $\Ma_c$  as a function of the Grashof number $G$. Right (b): critical wavenumber 
corresponding to the analysis in the top panel.}
\label{fig:ma:stab:2} 
\end{figure}
%-------------------------------------------------------

According to the critical value $\Ma^I_c$, two  nonlinear simulation are performed at \tk{$\Ma^I=-76$} (LS01) and \tk{$\Ma^I=-77$} (LS02), 
respectively (see Tab.~\ref{tab:nondim-parameters:dns}).  Their root mean square (rms) velocity which is defined by
\beq
u^{(2)}_{rms}(t)=\sqrt{\langle \mbf u(t)^2\rangle_{\Omega^{(2)}}}\,, 
\label{rmsvel}
\eeq
is plotted in the bottom panel of Fig.~\ref{fig:ma:stab:1}
as function of time, successfully verifying the prediction from the eigenvalue calculation.  In Fig.~\ref{fig:ma:stab:1a}, the 
eigenfunctions (crosses) at the critical parameters $k_c,\Ma_c^I$ are plotted together with the respective simulation profiles (solid lines).
Pointwise differences between two snapshots are compared. Eigenfunctions have been rescaled to have a range of values which
is comparable with the simulation results. \tk{Convection is caused at both interfaces since the temperature perturbation amplitude 
is non-zero there. At a given horizontal position, the temperature may be locally increased at the upper interface ($\theta(z=d_{21})>0$) 
causing a lower interfacial tension which drives a divergent flow at this point, i.e., a flow directed towards the interface with 
$\partial_z w(d_{21})< 0$. At the same horizontal position,  interfacial tension is then increased on the lower interface, and the flow 
is directed \tk{away from} the interface.}  At both interfaces work is performed on the liquid metal phases. 
%-------------------------------------------------------
\begin{figure}
\includegraphics[width=0.45\textwidth]{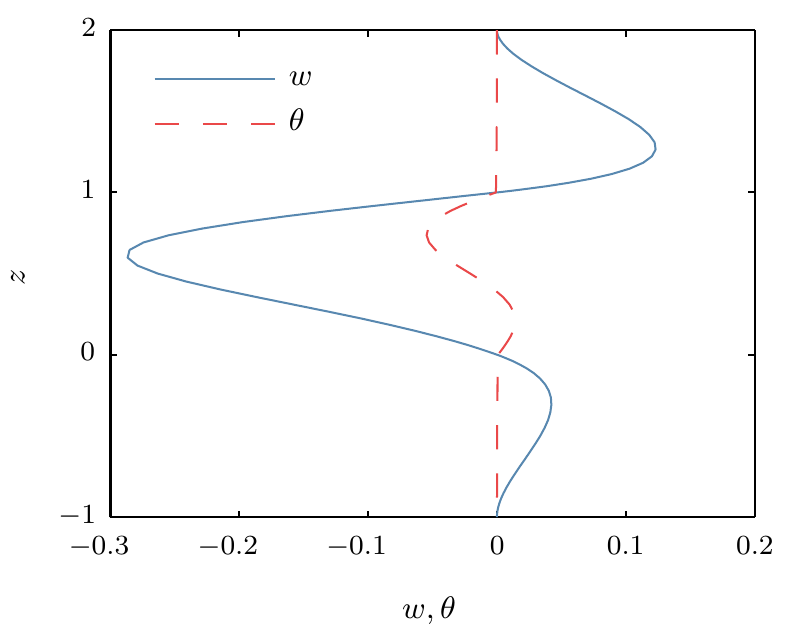} 
\caption{Eigenfunction profiles across the full battery height for pure Rayleigh-B\'{e}nard convection at 
$\Ma=0$, $\tilde G_c=-1.2937\times 10^4$ and $k_c=3$, respectively.}
\label{fig:ma:stab:2a} 
\end{figure}
%-------------------------------------------------------

However, the  upper interface II induces a stronger flow than the lower interface I as seen in the bottom panel of Fig.~\ref{fig:ma:stab:1a}. 
To understand the impact of each interface separately, we calculated the critical Marangoni number for each interface by disabling the effect of the 
other. When the \emph{lower} interface is disabled, i.e., $\Ma^I=0$, the value of \tk{$\Ma^{II}_c=-233.55$} is obtained which corresponds 
with \tk{$\Ma^{II}_c/\xi =-90.87$}, at a critical wavenumber $k_c=2.2$ which is very close to the value for both interfaces being active.
When the upper interface is disabled, i.e.,  $\Ma^{II}=0$, one obtains $\Ma^I_c=-106.88$ at $k_c=2.25$. The latter case of $\Ma^{II}=0$, is 
also verified by two DNS, one at $\Ma^I=-105$, showing decay and $\Ma^{II}=-110$, showing a growth of perturbations. Consequently, 
both interfaces collectively contribute to the instability. Moreover, the eigenfunctions with one active interface (not displayed) look similar 
to the eigenfunctions that result from two active interfaces. 
%-------------------------------------------------------
\begin{table*}
\begin{tabular}{c c  c c c c c c c c c c c}
\hline\hline\noalign{\smallskip}
Case& $d_{21}$ & $G$ & $\xi$ &$Ma$ & $l_x$ & $N_x$& $N_y$ &  $N_z^{(1)}$ & $N_z^{(2)}$& $N_z^{(3)}$\\
\hline
LS01  & 1&     0 & 2.57 &  \tk{-76} & 12 & 128& 128 & 32 & 64 &  32  \\  %varMa\_1a
LS02  & 1&    0 & 2.57 & \tk{-77} & 12 & 128 & 128 & 32 & 64 &  32 \\  %% varMa\_1b
LS03  & 1&  -1.29$\times 10^4$ & 0 & 0& 12 & 128 & 128 & 32 & 64 &  32 \\   %%varMa\_11
LS04  & 1&  -1.30$\times 10^4$ & 0 & 0 & 12 & 128& 128 & 32 & 64 &  32 \\  %%varMa\_12
LS05 & 0.3&  0 & 2.57 & \tk{-229} & 12 & 128& 128 & 32 & 64 &  32 \\  %%varMa\_17
LS06 & 0.3&  0 & 2.57 & \tk{-230} & 12 & 128 & 128 & 32 & 64 &  32 \\  %%varMa\_18
LS07 & 0.3&  -3.75$\times 10^5$ & 0 & 0 & 6 & 128& 128 & 32 & 64 &  32 \\ %%%   varMa\_19
LS08 & 0.3&  -3.85$\times 10^5$ & 0 & 0 & 6 & 128& 128 & 32 & 64 &  32 \\  %%%  varMa\_20
LS09 & 0.1&  -2.30$\times 10^6$ & 0 & 0 & 6 & 128& 128 & 32 & 64 &  32 \\  %%%varMa\_21
LS10 & 0.1&  -2.40$\times 10^6$ & 0 & 0 & 6 & 128& 128 & 32 & 64 &  32 \\%%%% varMa\_22
LS11 &  1& -1.3314$\times 10^4$ & 2.57& -10.16 & 6 &128& 128 & 32 & 64 &  32  \\  %%%% sb\_4
LS12 &  1& -1.3314$\times 10^4$  & 0   &    0    & 6 &128& 128 & 32 & 64 &  32  \\  %%% sb\_4\_noMa
\hline\hline
\end{tabular}
\caption{List of full nonlinear DNS runs to verify the eigenvalues and eigenvectors of the numerical solution
of the linear stability analysis. Grashof and Marangoni numbers as well as spectral resolutions are listed.  
All other parameters not listed here correspond with those listed in Tab.\ref{tab:nondim-parameters:ref}.}
\label{tab:nondim-parameters:dns}
\end{table*}
%-------------------------------------------------------

\subsubsection{Rayleigh-Marangoni instability ($\Ma\ne 0$, $G\ne 0$)} 
\label{sec:stab:RMC}
Next, a closer look at the impact of buoyancy effects with $G\ne 0$ is taken.  In order to do so, we calculate the 
critical Marangoni number $\Ma^I_c$ as a function of the Grashof number $G$. Figure.~\ref{fig:ma:stab:2}(a) 
shows this computed relation as crosses, while Fig.~\ref{fig:ma:stab:2}(b) displays the corresponding critical 
wavenumber. The magnitude of the critical Marangoni number decreases as |$G$| is increased.  
In what follows, the relation $\Ma^{II}=\xi \Ma^I$ will always hold. Thus, we will continue with the single Marangoni 
number $\Ma \equiv \Ma^I$ only. An approximately linear trend of the form,
%-------------------------------------------------------
\beq
\frac{\Ma_c}{\tilde \Ma_c} = 1- G/\tilde G_c, \label{eq:threshold:RMC}
\eeq
%-------------------------------------------------------
is observed, where $\tilde \Ma_c$ and  $\tilde G_c$ are the critical numbers in absence of the correspondingly other
physical process. Such linear scaling has been already observed in the one-layer case \citep{nield1964surface} when both, 
the Marangoni and the Rayleigh instability, destabilize the system.

The particular value of the critical Grashof number $\tilde G_c$ is calculated from an eigenvalue problem similar to 
\eqref{eq:eigenvalue:problem:1} but now for Grashof instead of Marangoni number. We find a critical value 
$\tilde G_c=-1.2937\times 10^4$ with a critical wavenumber of $k_c=3$. The eigenfunctions of the pure Rayleigh-B\'{e}nard 
case are displayed in Fig.~\ref{fig:ma:stab:2a}. They show a similar flow structure as in the Marangoni case, though now, 
caused by the downwelling of colder fluid in the upper half of the mid layer. The flow in the top layer is caused by viscous friction 
at the interface. In this layer, temperature perturbations are negative which result in a stabilization. 

Again we compared these findings to two DNS simulations, \tk{namely DNS runs LS03  and LS04. After a decay of initial perturbations, 
the unstable modes start to develop in the supercritical case LS04 while perturbations decays in the subcritical case LS03.}

The middle layer Rayleigh-B\'{e}nard mode leads to  eigenfunctions that are qualitatively equal to those of Marangoni convection. As a 
consequence, Rayleigh-B\'{e}nard (RC) and Marangoni convection (MC) act together to destabilize the system at hand. The case when 
both sources for convection are present will be denoted by Rayleigh-Marangoni convection (RMC). The Rayleigh and Marangoni modes 
act together in the middle layer of the present three-layer system as in the single layer case. The actual thresholds can be calculated by 
Eq.~\eqref{eq:threshold:RMC}.  

%-------------------------------------------------------
\begin{figure}
\includegraphics[width=0.4\textwidth]{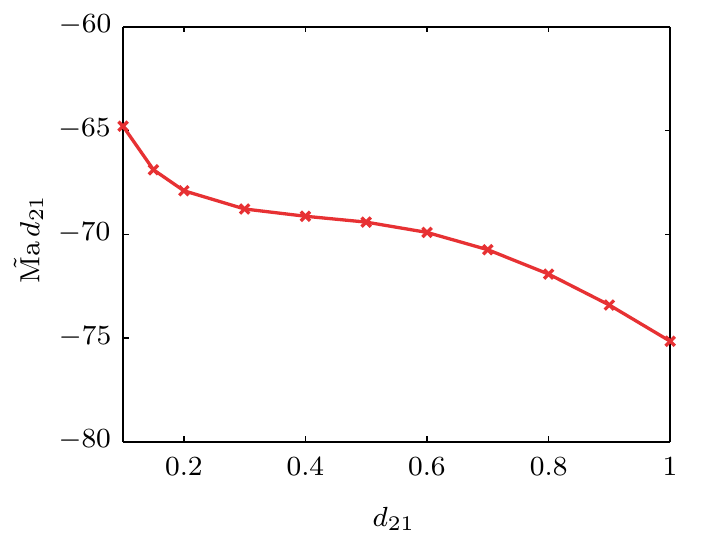}
\caption{Linear stability analysis of \tk{Marangoni convection (MC)} for varying middle layer heights 
$d_{21}$. The compensated critical Marangoni number $\tilde \Ma_c\, d_{21}$  as a function of the middle layer height 
$d_{21}$ is displayed.}
\label{fig:ma:stab:3a} 
\end{figure}
%-------------------------------------------------------
\begin{figure}
\includegraphics[ width=0.4\textwidth  ]{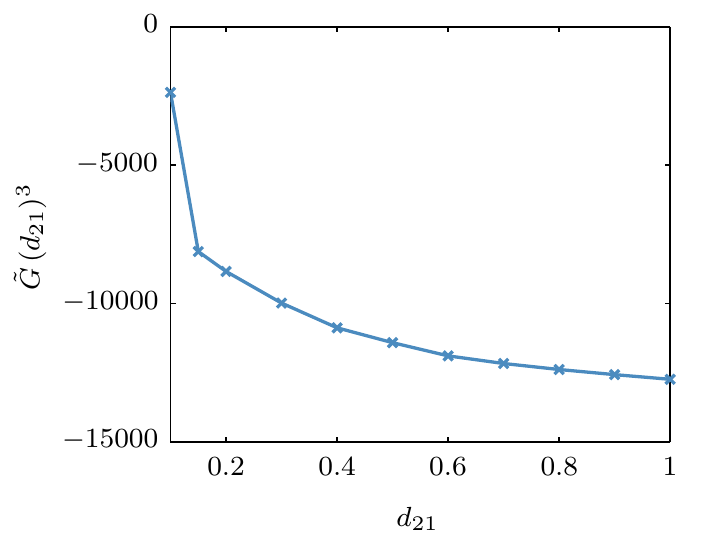}
\caption{Plot of the compensated critical Grashof number $\tilde G_c\,(d_{21})^3$ as a function of the middle layer height $d_{21}$.}
\label{fig:ma:stab:3c} 
\end{figure}
%-------------------------------------------------------
\subsubsection{Middle layer thickness}  
While a change in the electrical current density magnitude and the overall system height at fixed ratios $d_{21}$ and $d_{31}$ 
is accounted for in the Marangoni and the Grashof numbers, respectively, the stability threshold is changed when the mid-layer 
thickness $d_{21}$ becomes smaller compared to the outer layer, i.e., $d_{21}<1$ in our notation. Therefore, we also computed 
the stability thresholds for both convection mechanisms as a function of the middle layer height $d_{21}$.

Figure~\ref{fig:ma:stab:3a} shows how the magnitude of $\tilde \Ma_c$ grows with decreasing layer height, which is depicted  
with the compensated plot of $\tilde \Ma_c\,d_{21}$ as a function of $d_{21}$. In this case, $G=0$ was taken. For example, the threshold 
values for $d_{21}=0.3$ are \tk{$\Ma_c=229.2$} and $k_c=6.3$. For this case, we verified again by two DNS, \tk{ namely, LS05 and LS06}, 
that this Marangoni number is indeed critical. 
%-------------------------------------------------------
\begin{figure}
\includegraphics[width=0.38\textwidth]{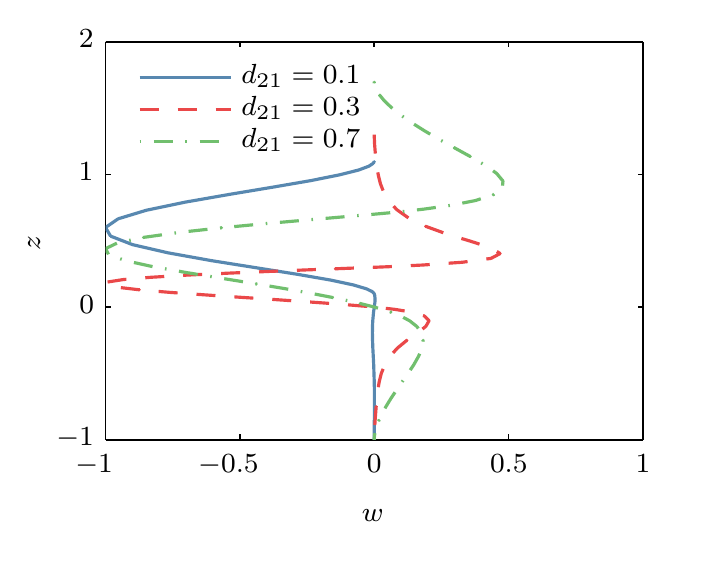}
\caption{Eigenfunctions at $\Ma=0$ of velocity perturbation $w(z)$ for three mid-layer heights corresponding to the critical parameters.
The relative thickness of the middle layer height is indicated in the legend.}
\label{fig:ma:stab:3d} 
\end{figure}
%-------------------------------------------------------

The magnitude of the critical Grashof number ($\Ma=0$) increases as well with decreasing middle layer height. The critical Grashof number 
which is compensated with the third power of $d_{21}$ is shown in Fig.~\ref{fig:ma:stab:3c}. In this compensated representation, one observes 
that the Grashof number increases less than the third power of $d_{21}$. For the shallowest layer $d_{21}=0.1$ one observes a particularly 
strong decrease of $\tilde G_c\, (d_{21})^3$. 

An inspection of the corresponding eigenfunctions reveals how the velocity perturbation changes with $d_{21}$. The results are shown in 
Fig.~\ref{fig:ma:stab:3d}.  For a shallow middle layer with $d_{21}=0.3$, the upper layer, which is initially coupled by viscous shear only,  
gets active and develops \tk{internal convection rolls.} This is because the coupling of layers by viscous stresses and heat transport are 
not mutually  'compatible' due to the much higher thermal diffusivity in the liquid metal top layer.  For $d_{21}=0.1$, convection in the 
upper-layer layer dominates.  A corresponding upper layer Rayleigh number can be calculated by  
%-------------------------------------------------------
\beq
\Ra^{(3)}=\frac{G \Pr T^{II} {d_{31}}^3}{ \kappa_{31}\nu_{31}\beta_{T,31}}=\frac{g \beta_T^{(3)} \left(d^{(3)}\right)^3 ({\tilde{T}}^{II}-T^\infty )}{\nu^{(3)}\kappa^{(3)}}. 
\label{eq:upperlayer:Rayleigh}
\eeq
%-------------------------------------------------------
with $T^{II}=(\tilde{T}^{II}-T^\infty)/\Theta$  (see Eq.~\eqref{Grashof}).
For the calculated threshold ($d_{21}=0.1$) of $G_c=-2.37\times 10^{6}$, $k_c=2.55$, this number equals to $\Ra^{(3)}=-1290.4$. This 
should be compared to the classical linear stability analysis in one layer where the critical Rayleigh number ranges from $\Ra_c=120$ with 
adiabatic temperature ($D\theta=0$) and free-slip velocity ($w=D^2w=0$) boundary conditions to a value of $\Ra_c=1707.76$ with 
isothermal temperature ($\theta=0$) and no-slip velocity ($w=Dw=0$) boundary conditions \citep{colinet2001nonlinear}. Hence for shallow 
middle layers a fourth regime of convection can be expected that is similar to the classical Rayleigh-B\'{e}nard convection, caused by the linear 
temperature gradient in the upper layer. Note also that we probed successfully the stability threshold for $d_{21}=0.1, 0.3$ with DNS simulation 
runs LS07 to LS10 (see again Tab.~\ref{tab:nondim-parameters:dns}). 

Finally, it is investigated how both effects work together, in the same way as in the previous Sec.~\ref{sec:stab:RMC}. It is found that relation 
\eqref{eq:threshold:RMC} can be applied up to $d_{21}\ge 0.2$. However, for $d_{21}=0.1$, we found two different behaviors of $G_c(\Ma)$: 
first, for \tk{$|\Ma| < 400$}, the Marangoni effect counteracts the upper layer RC, such that the critical Grashof number grows in magnitude with 
increasing $|\Ma|$. Second, for |Ma|>\tk{400}, the middle layer RC is dominant, thus, $|G_c|$ is decreasing with increasing $|\Ma|$.

%-------------------------------------------------------
\begin{figure}
\includegraphics[ width=0.4\textwidth]{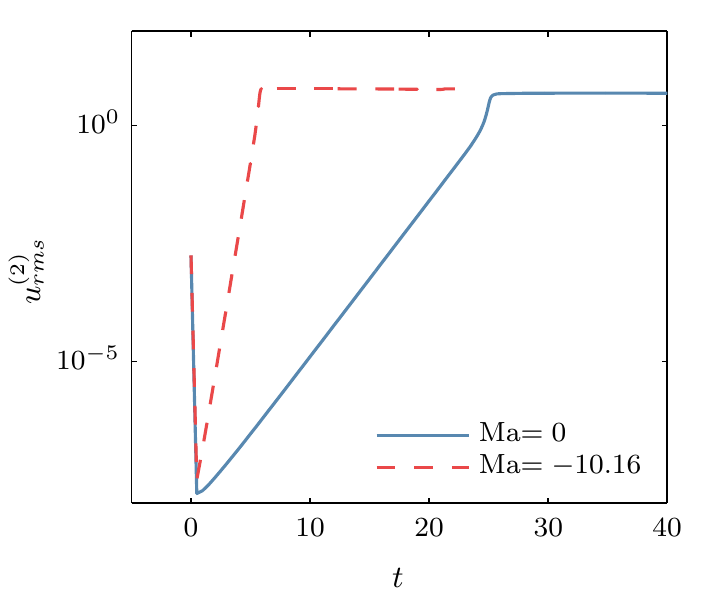}
\caption{Simulation of convection near the threshold of $G=-1.33\times10^4$ for the material parameters derived in Tab~\ref{tab:physprop:ref} 
but a shallower layer of $d^{(i)}=6.4$ mm (simulation runs LS11 and LS12). Root mean square velocity in the middle layer as function of time
is displayed.} 
\label{fig:case:sb4a} 
\end{figure}
%-------------------------------------------------------
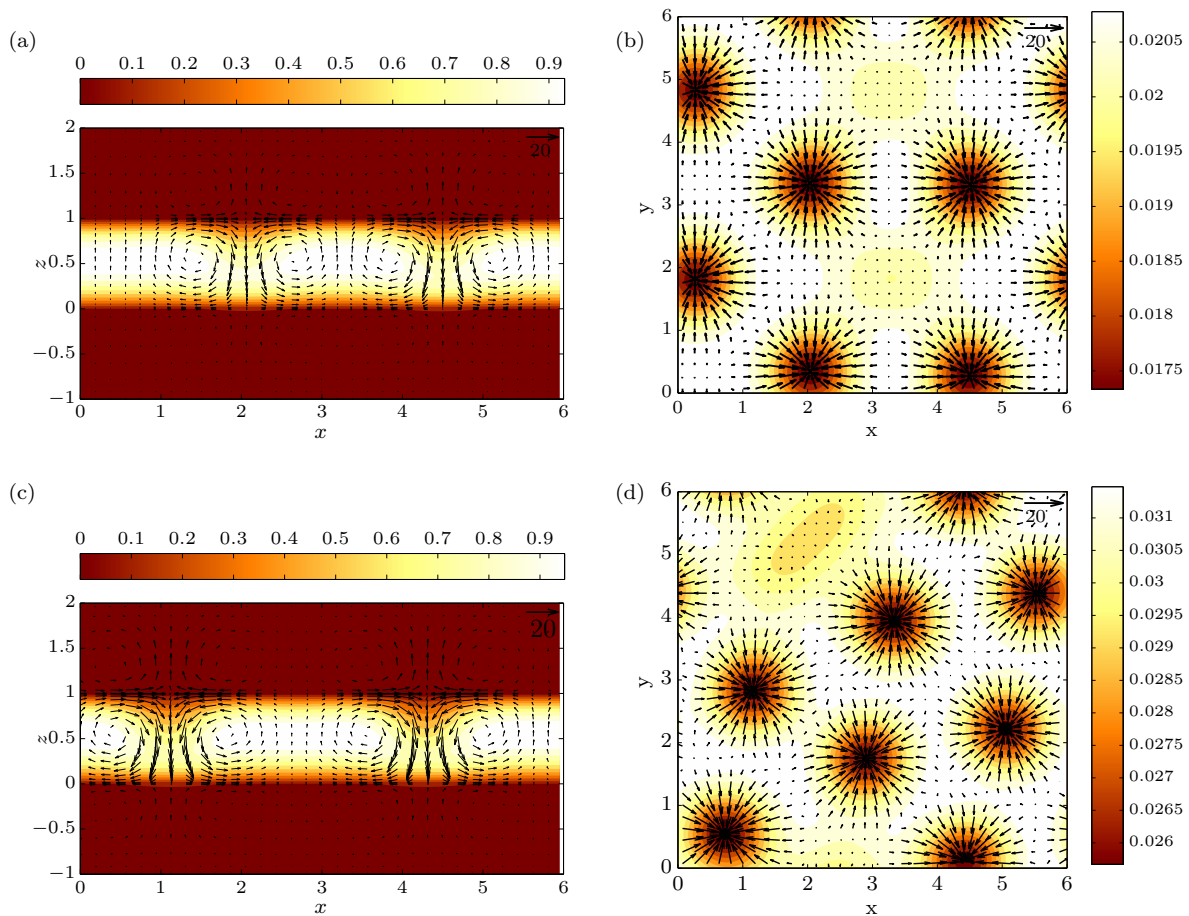
\begin{figure*}
	\centering
		\begin{tabular}{cc}
\includegraphics[ width=0.45\textwidth]{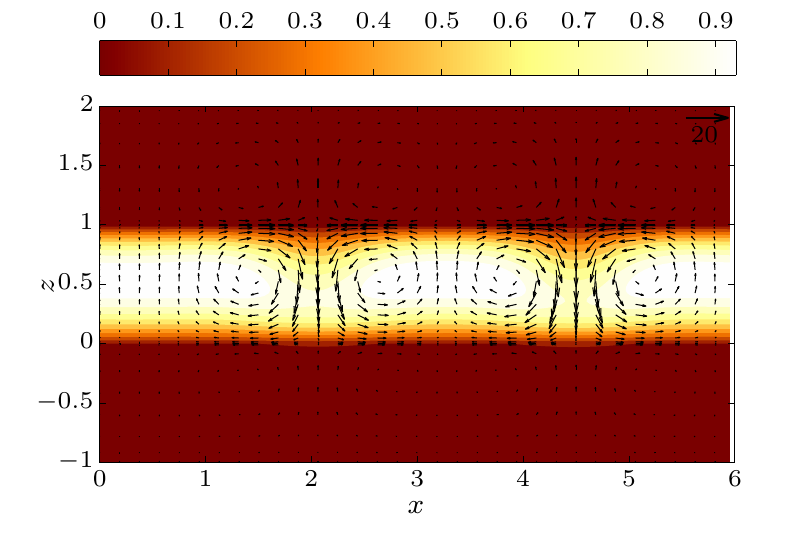} &
\includegraphics[ width=0.45\textwidth]{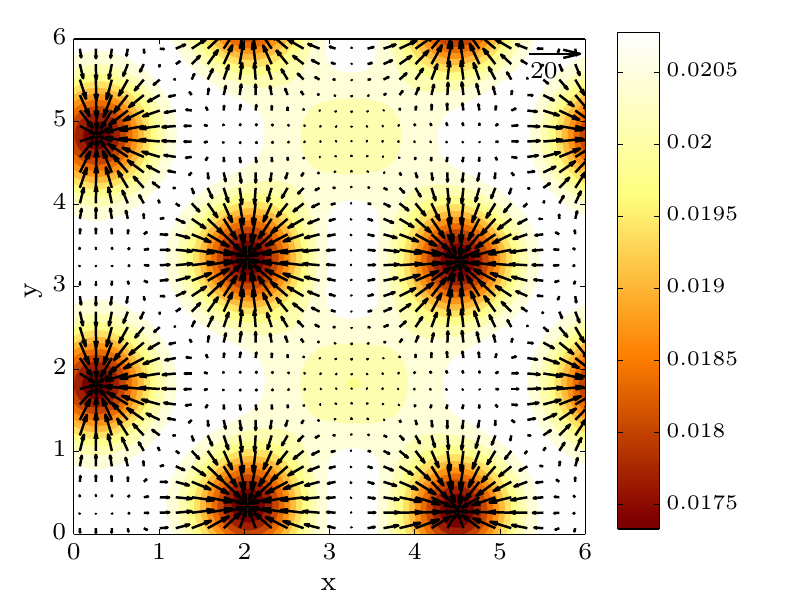} \\

\includegraphics[  width=0.45\textwidth]{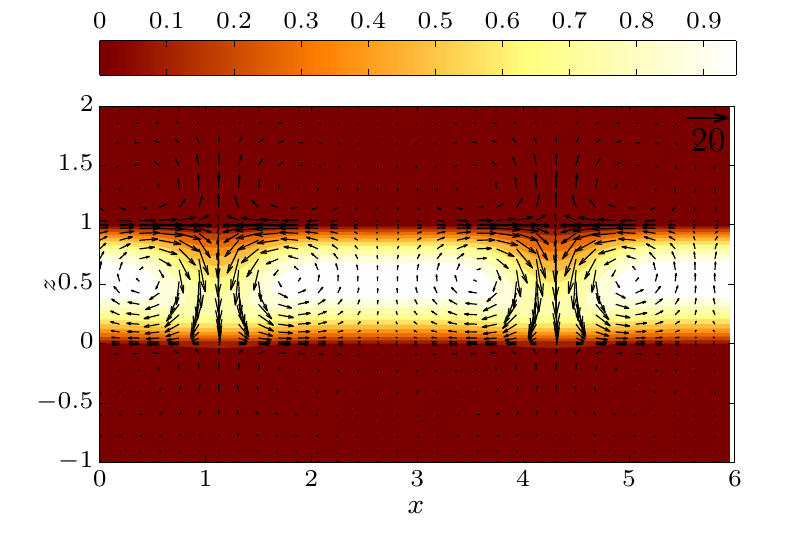} &
\includegraphics[  width=0.45\textwidth]{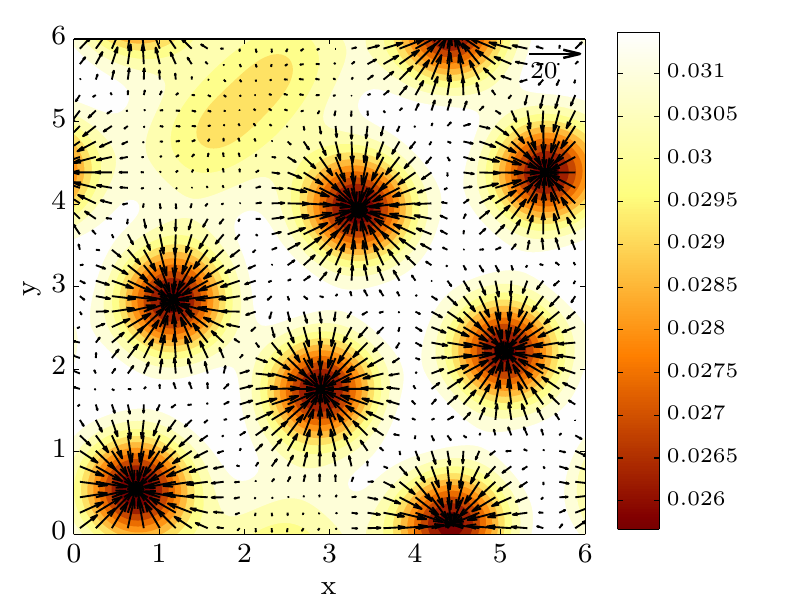} \\
\setlength{\unitlength}{0.5\textwidth}
\begin{picture}( 0, 0)(0,0)
\put(-0.45 , 1.35){\footnotesize (a)}
\put(-0.45 , 0.68){\footnotesize (c)}
\put(0.45 , 1.35){\footnotesize (b)}
\put(0.45 , 0.68){\footnotesize (d)}
\end{picture}
	\end{tabular}
\caption{Contour-vector plots of temperature and velocity of the simulations of convection near the threshold. 
(a) Simulation snapshot of LS12 with $\Ma=0$ at $t=30$ shows a vertical cut at $y=0$ with temperature contours and velocity field 
vectors. (b) Velocity and temperature for the same data at the upper interface II. (c,d) the same quantities as (a,b) for LS11 at $t=20$.}
\label{fig:case:sb4} 
\end{figure*}
%-------------------------------------------------------

\subsubsection{Convection patterns near the threshold values} 
The  characteristic structures of convection near the onset threshold are shown in two DNS which are denoted LS11 and LS12, respectively. 
Here, we used the reference current $j_0=3$kA/m$^2$, but adapted the height to the threshold of the onset of Rayleigh-B\'{e}nard convection. 
Therefore, all three layer heights are set to $d^{(1)}=6.4$ mm, which results in an overcritical Grashof number (see DNS run LS11 in 
Tab.~\ref{tab:nondim-parameters:dns}).  Moreover, a simulation LS12 with $\Ma=0$, but otherwise the same parameters as LS11 was
conducted.

The rms velocity $u_{rms}^{(2)}(t) $ for the two simulations are shown in Fig.~\ref{fig:case:sb4a}. The solid line represents Rayleigh-B\'{e}nard
convection and the dashed line Rayleigh-Marangoni convection. We observe that the Marangoni effect accelerates the growth of the rms
velocity amplitude and thus of convection. The larger growth rate is expected from our former results. The flow structures of these simulations  
are visualized in Fig.~\ref{fig:case:sb4}(a-d). We show vertical cuts at $y=0$ with the temperature and  velocity for RC in panel (a) of the figure 
and RMC in panel \tk{(c)}. Furthermore, the interfacial temperature and 
velocity vectors at the same time instant are displayed in panels \tk{(b,d)}. A nearly hexagonal and stationary convection  pattern appears with 
an outflow in the center of cells. This is similar to observations for one layer with internal convection \cite{kulacki1972thermal,tveitereid1976convection}. 
In conclusion, the former results that the Rayleigh and Marangoni effect act together in the exemplary configuration are hereby again confirmed.

\section{Nonlinear evolution}
\label{sec:nonlinear}

%--------------------------------------------------------------------
\begin{table*}
\begin{tabular}{l c c c c c c c c c c c c c c c c c c}
\hline\noalign{\smallskip}
\hline\
Case &  $d_{21}$ &  $G$ & $\tilde G_c$ & $k^{\rm{RC}}_c$ & $\Ma$ & $\tilde \Ma_c$ & $k^{\rm{MC}}_c$ & $N_x (=N_y)$ & $l_x (=l_y)$ &  $N_z^{(1)}$ & $N_z^{(2)}$ & $N_z^{(3)}$ & $d^{(1)}$ [mm] & 
$j_0$ [kA/m$^2$]\\
\hline
H1$^+$ & 1 & -3.88$\times 10^3$ & -1.29$\times 10^4$ & 3   & -4.84 & -\tk{76.4} & \tk{2.10} &  128  & 6 & 32 & 64 & 32  &  5 &3\\
H2 & 1 & -1.24$\times 10^5$ &  -1.29$\times 10^4$ & 3  & -38.75 & -\tk{76.4} & \tk{2.10} &  128 &6 &  32 & 64 & 32 &  10 &3\\
H3 &  1  & -9.42$\times 10^5$ & -1.29$\times 10^4$ &3  & -130.79& -\tk{76.4} & \tk{2.10}  &128 & 6 &  32 & 64 & 32   &15& 3\\
H4 &  1  & -3.97$\times 10^6$ & -1.29$\times 10^4$ &3  & -310.02 & -\tk{76.4} & \tk{2.10} & 256 & 6 &  32 & 64 & 32   &20 &3\\
H5 &  1 & -3.01$\times 10^7$ & -1.29$\times 10^4$  &3  & -1.05e3 & -\tk{76.4} & \tk{2.10}   & 256 & 6 &  64& 128 & 64   & 30&  3\\
H6 &  1  & -1.27$\times 10^8$ & -1.29$\times 10^4$ &3  & -2.48e3 & -\tk{76.4} & \tk{2.10}   &  512 & 6 &  64 & \tk{128} & \tk{64} &  40&  3\\ \\

JM1 & 0.5& -1.10$\times 10^5$ & -9.23$\times 10^4$ &5.75 & -8.61    & \tk{-138.8} &\tk{3.75}   &  128 & 3 &  32 & 64 & 32& 20& 1 \\				
JM2 & 0.5& -9.92$\times 10^5$  & -9.23$\times 10^4$ &5.75 &  -77.50 & \tk{-138.8} & \tk{3.75}   &  128 & 3 &  32 & 64 &32 & 20& 3 \\				
JM3 & 0.5& -2.76$\times 10^6$ & -9.23$\times 10^4$ &5.75 &  -215.29  & \tk{-138.8} & \tk{3.75}  & 256 & 3 &  64 & 128 & 64 & 20& 5 \\				
JM4 & 0.5& -1.10$\times 10^7$ & -9.23$\times 10^4$ &5.75 &  -861.16 & \tk{-138.8} &\tk{3.75}   & 256 &3 &  64 &128 & 64 & 20& 10 \\ \\			

JS1$^+$ & 0.3& -3.97$\times 10^4$ & -3.81$\times 10^5$ &9.30&   -3.1 & \tk{-229.2} & \tk{6.25}   	& 128 & 2 & 32 & 32 & 32 & 20& 1 \\				
JS2$^+$ & 0.3& -3.57$\times 10^5$ & -3.81$\times 10^5$ & 9.30&   -27.91  & \tk{-229.2} & \tk{6.25}   & 128 & 2 & 32 & 32 & 32 & 20& 3 \\				
JS3 & 0.3& -9.92$\times 10^5$ & -3.81$\times 10^5$ & 9.30&   -77.50 & \tk{-229.2} & \tk{6.25}    &128 & 2 &  32 & 32 & 32 & 20& 5 \\				
JS4 & 0.3& -3.97$\times 10^6$ & -3.81$\times 10^5$ & 9.30&    -310.01 & \tk{-229.2} & \tk{6.25}  & 256 & 2 &  64 & 64 & 64   & 20& 10 \\
JS5 & 0.3& -1.59$\times 10^7$ & -3.81$\times 10^5$ & 9.30&   -1240.1  & \tk{-229.2} & \tk{6.25}  &  256 & 2 &  64 & 64 & 64   & 20& 20 \\ \\				
			
JXS1$^+$ & 0.1& -1.76$\times 10^6$ &  -2.37$\times 10^6$ & 2.55& -137.78 & \tk{-647.9}& \tk{19.25} & 128 & 6 & 64 & 32 & 64 & 20& 20 \\				
JXS2 & 0.1& -4.81$\times 10^6$ &  -2.37$\times 10^6$ & 2.55&  -24.80 & \tk{-647.9} & \tk{19.25} & 128 & 6 & 64 & 32 & 64 & 40& 3 \\				
\hline 
\noalign{\smallskip}\hline
\end{tabular}
\caption{List of full nonlinear simulation runs with different electrical current densities $j_0$ and/or layer heights $d^{(1)}$.   
For each case displayed, an additional simulation with the Marangoni effect disabled is performed. Again $N_x=N_y$ and 
$l_x=l_y$. The runs with superscript ``+ '' are found to be linearly stable.}
\label{tab:paramteric-study:height}
\end{table*}
%--------------------------------------------------------------------

In the following section we will present a series of DNS, listed in Tab.~\ref{tab:paramteric-study:height}, that investigate 
the full nonlinear and turbulent evolution of the RMC in the three-layer battery model. Several parameters 
will be varied for this purpose. Runs H1 to H6 are conducted at $d_{21}=1$ and fixed $j_0$, but different dimensional heights 
$d^{(1)}$. An increase of $d^{(1)}$ is in line with an increasing Grashof number. In the series JM1 to JM4, we fix the 
physical height of the bottom layer $d^{(1)}=20$ mm as well as $d_{21}=0.5$ while varying $j_0$ (J = varying current density, 
M = medium middle-layer height). The same holds for the series JS1 to JS5 where $d_{21}$ is further decreased to 0.3 in 
comparison to the JM series (S = small).  Finally, the runs JXS1 and JXS2 set $d_{21}$ to 0.1 (XS = extra small).   
%----------------------------------------------------
\begin{figure}
\includegraphics[width=0.38\textwidth]{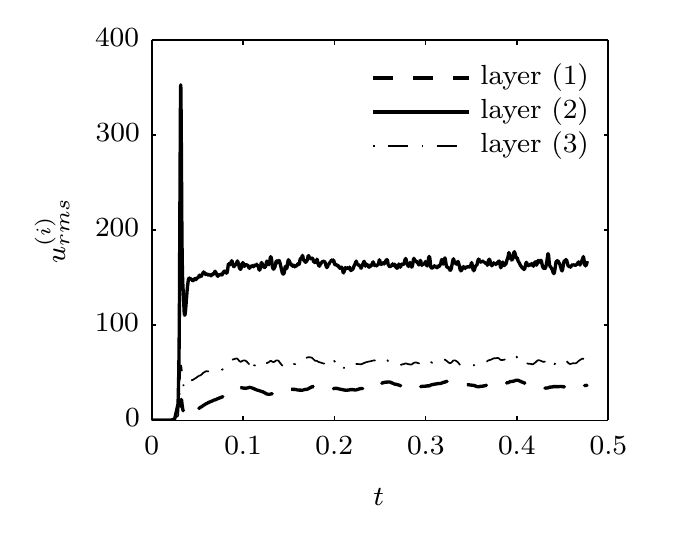} \\
\includegraphics[width=0.38\textwidth]{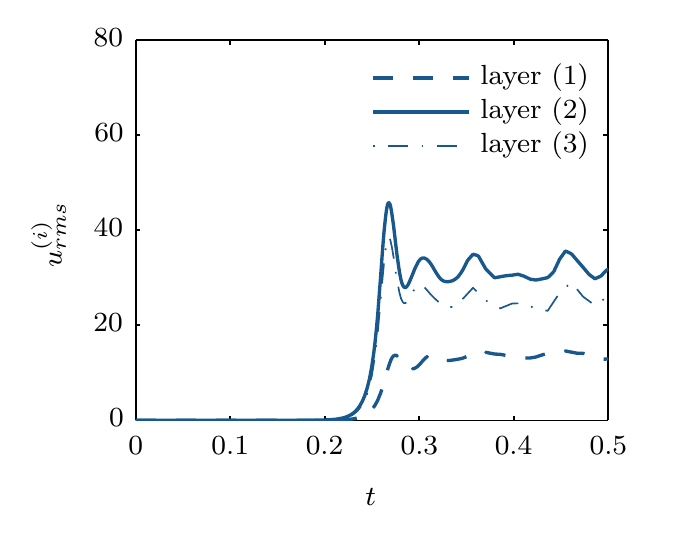} \\
\includegraphics[width=0.38\textwidth]{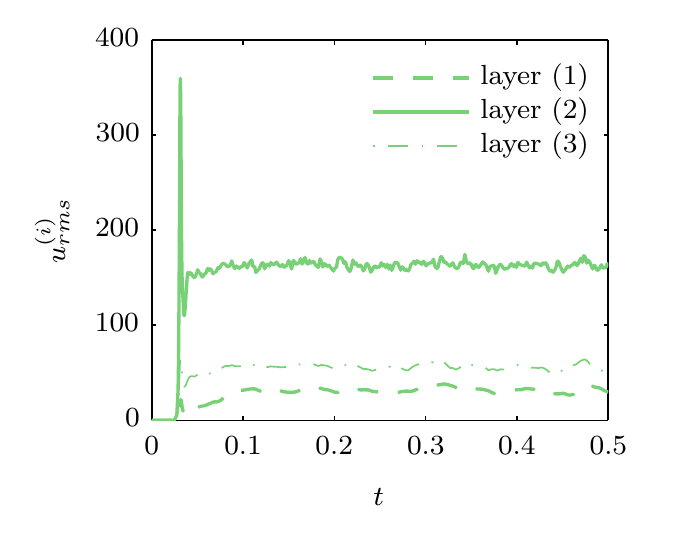} 

\setlength{\unitlength}{0.5\textwidth}
\begin{picture}( 0, 0)(0,0)
\put(-0.4 , 1.88){\footnotesize (a)}
\put(-0.4 , 1.25){\footnotesize (b)}
\put(-0.4 , 0.63){\footnotesize (c)}
\end{picture}
\caption{Convection in the reference configuration H4 of Tab.~\ref{tab:paramteric-study:height}.   
Root mean square velocities in each of the three layers versus time are shown for RC (a), MC 
(b) and RMC (c).} 
\label{fig:ref:integral} 
\end{figure}
%----------------------------------------------------
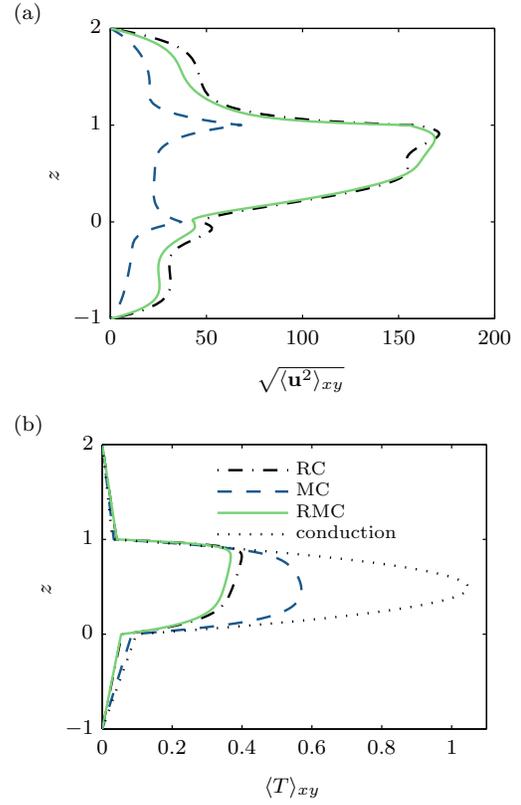
\begin{figure}
\includegraphics[width=0.38\textwidth]{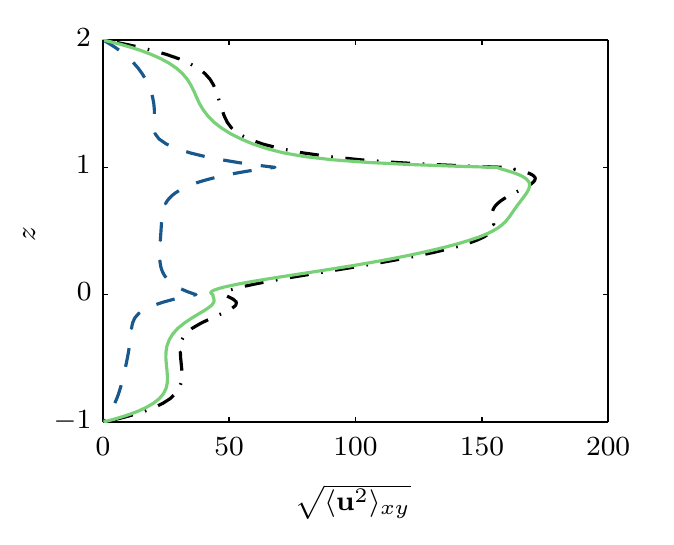} \\
\includegraphics[width=0.38\textwidth]{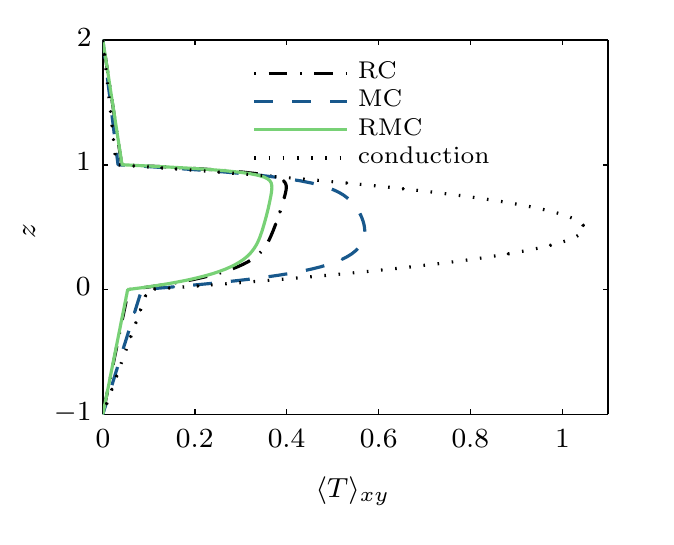} 
\setlength{\unitlength}{0.5\textwidth}
\begin{picture}( 0, 0)(0,0)
\put(-0.8 , 1.2){\footnotesize (a)}
\put(-0.8 , 0.59){\footnotesize (b)}
\end{picture}
\caption{Convection in the reference configuration H4 of Tab.~\ref{tab:paramteric-study:height}.   
Vertical profiles of the $x,y$-averaged rms velocity in cases RC, MC and RMC, respectively, are
shown in the top panel (a). $x,y$-averaged temperatures are shown in the bottom panel (b). All data are
also averaged in time. Furthermore, we plot the vertical temperature profile of the pure 
conduction state for comparison.} 
\label{fig:ref:integral1} 
\end{figure}
%----------------------------------------------------
\begin{figure}
\includegraphics[width=0.35\textwidth]{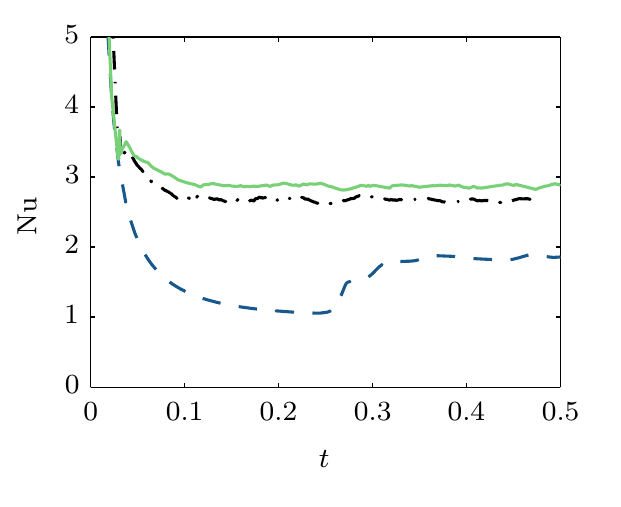} 
\caption{Convection in the reference configuration H4 of Tab.~\ref{tab:paramteric-study:height}.   
Nusselt number Nu as defined in Eq.~(\ref{Nuss_internal}) versus time for runs RC, MC and
RMC. Line styles agree with those in Fig.~\ref{fig:ref:integral1}. } 
\label{fig:ref:integral2} 
\end{figure}
%------------------------------------------------------

\subsection{Dynamics of the reference case H4}
\label{sec:res:ref}
As the reference case, we take the configuration with each layer having a height of 20 mm and a current density of 3kA/m$^2$, 
(see case H4 in Tab.~\ref{tab:paramteric-study:height}). All other parameters are given again in Tab.~\ref{tab:nondim-parameters:ref}. 
Although for commercial applications shallower middle layers are preferable to decrease Ohmic losses in the separator, a thicker 
middle layer helps us here to reveal flow properties in the cell center and at the interfaces. To separate the different convection 
mechanisms from each other, we performed three simulations (1) RC at $\Ma=0$, (2) MC at $G=0$, and (3) RMC. 
Table~\ref{tab:paramteric-study:height} also lists the critical Marangoni and Grashof numbers as well as the corresponding critical
wavenumbers which allow us to estimate the relative contribution of the effects. One finds a factor of supercriticality of $G/\tilde G_c=308$ 
for buoyancy--driven convection compared to \tk{$\Ma/\tilde \Ma_c=4$} for interfacial tension--driven convection.
%------------------------------------------------------
\begin{figure*}
\begin{tabular}{cc}
 \includegraphics[width=0.45\textwidth]{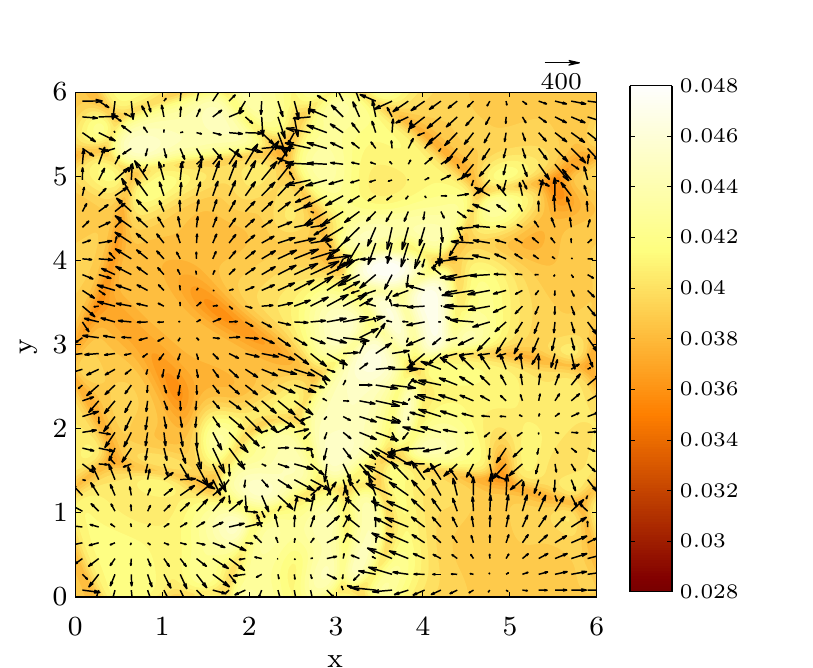}& 
 \includegraphics[ width=0.45\textwidth]{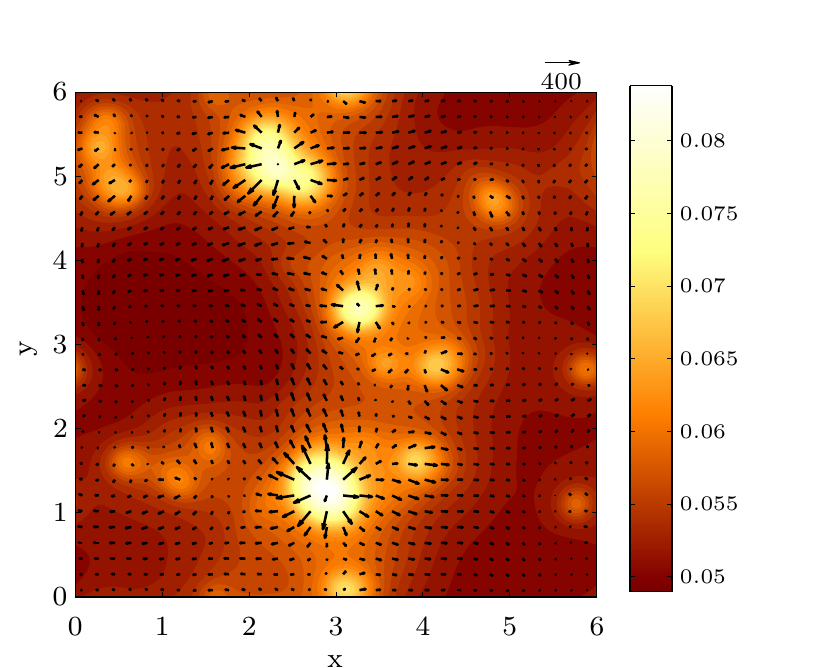} \\
 \includegraphics[ width=0.5\textwidth]{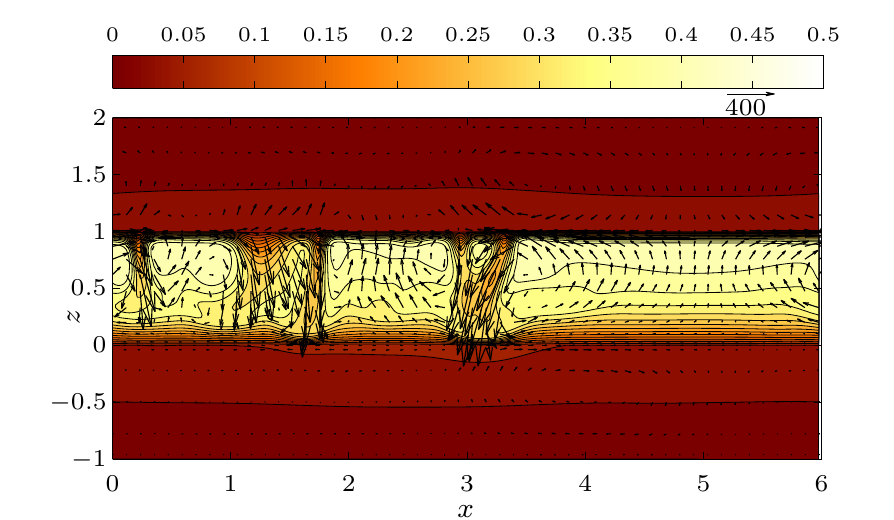} &
 \includegraphics[trim = 10px 10px 10px 10px, clip, width=0.35\textwidth]{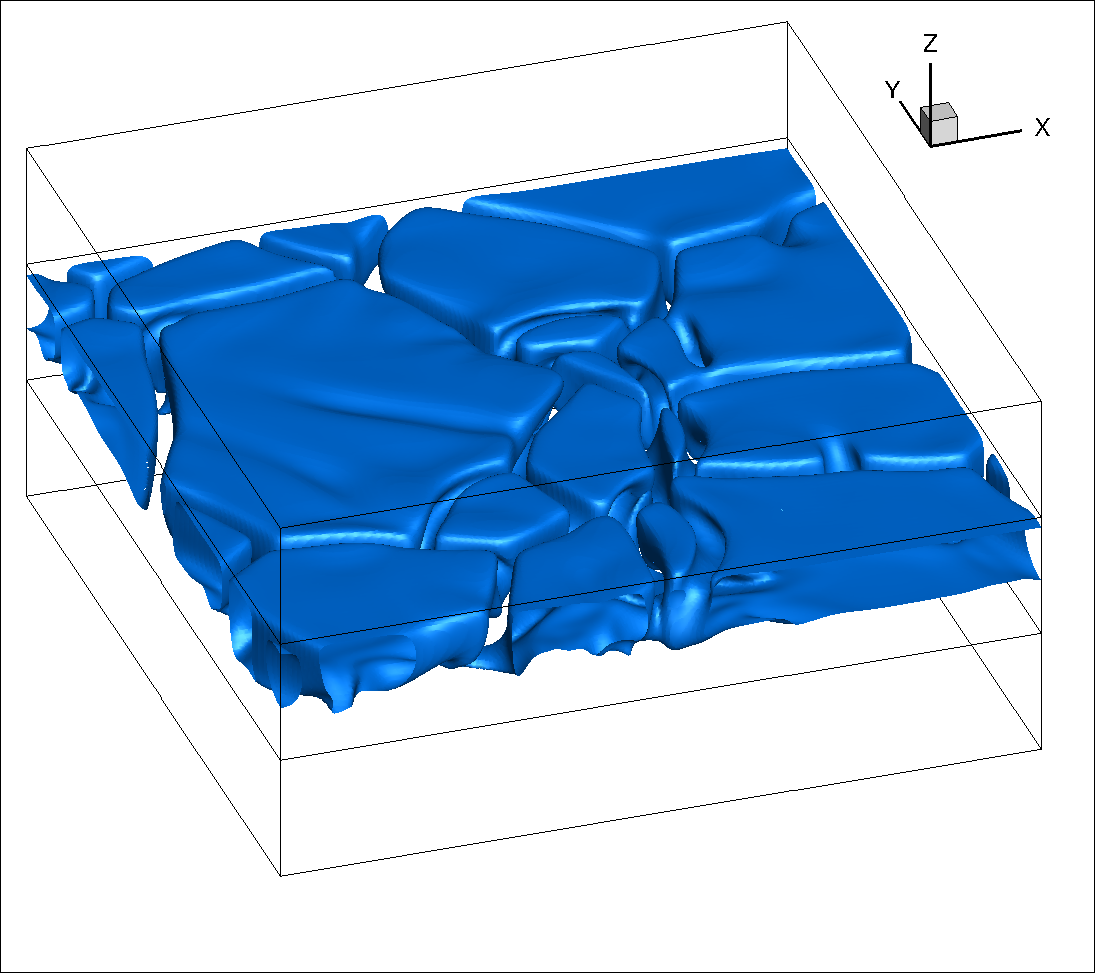}
 \setlength{\unitlength}{0.5\textwidth}
\begin{picture}( 0, 0)(0,0)
\put(-1.83 , 1.3){\footnotesize (a)}
\put(-0.85 , 1.3){\footnotesize (b)}
\put(-1.83 , 0.55){\footnotesize (c)}
\put(-0.85 , 0.55){\footnotesize (d)}
\end{picture}
\end{tabular}
\caption{Rayleigh-B\'{e}nard convection ($\Ma=0$) in the reference configuration H4. Temperature and velocity distributions at $t=0.3$
are shown. Temperature is displayed as a contour plot, velocity by arrows projected into the corresponding plane. (a) Temperature 
and velocity at the upper interface II. The length of the reference velocity arrows is always in units of $U_{vis}$. (b) Temperature and 
velocity at the lower interface I. (c) Vertical cut at $y=0$. (d) Isosurface of temperature at  $T=0.37$.}
\label{fig:pattterns:h4noMa} 
\end{figure*}
%-----------------------------------------------------
\begin{figure*}
\begin{tabular}{cc}
\includegraphics[width=0.5\textwidth]{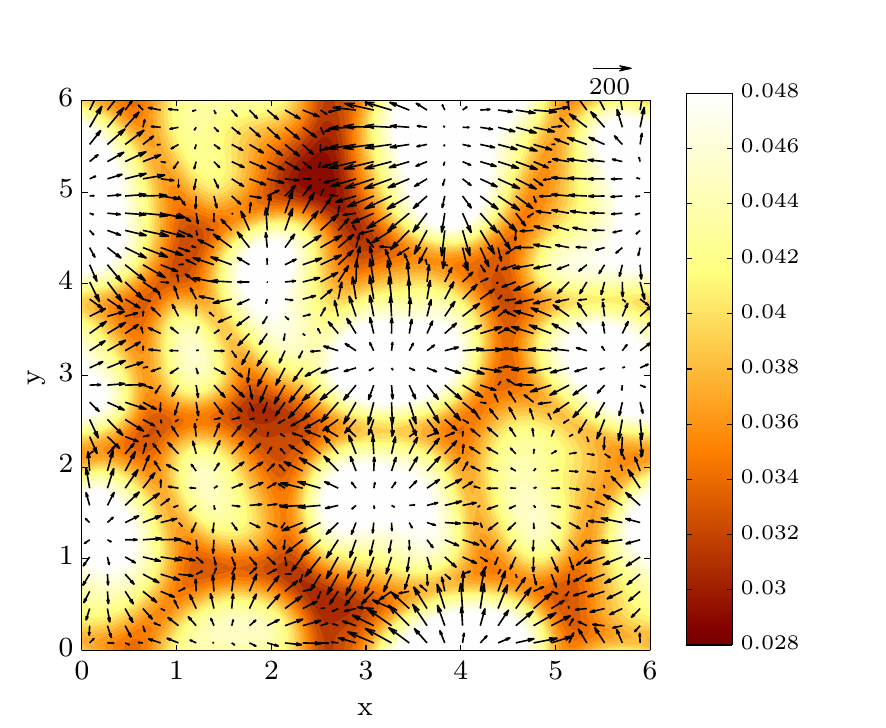}  & 
\includegraphics[ width=0.5\textwidth]{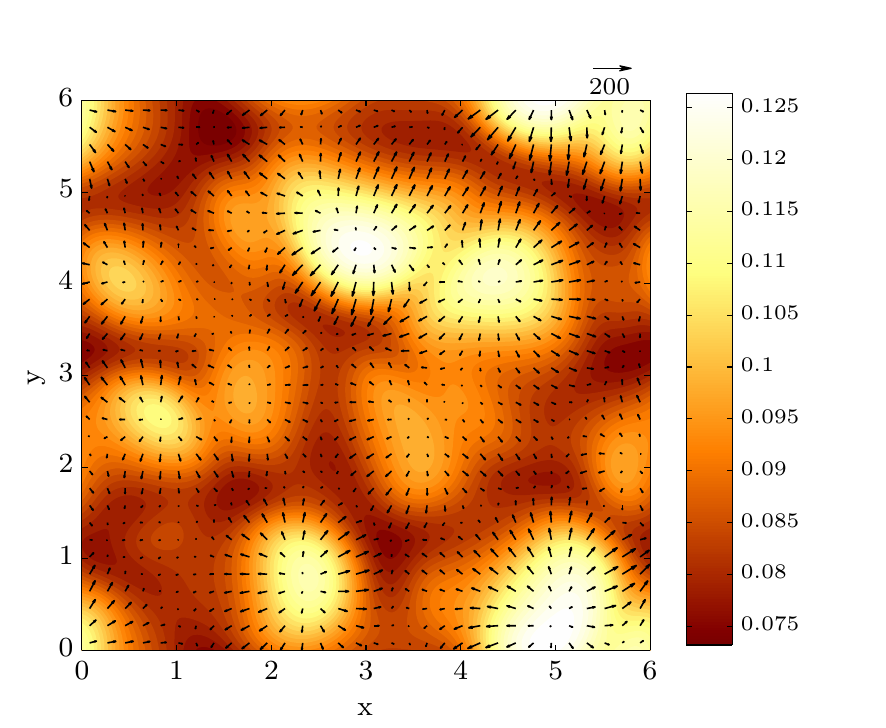}\\
\includegraphics[ width=0.55\textwidth]{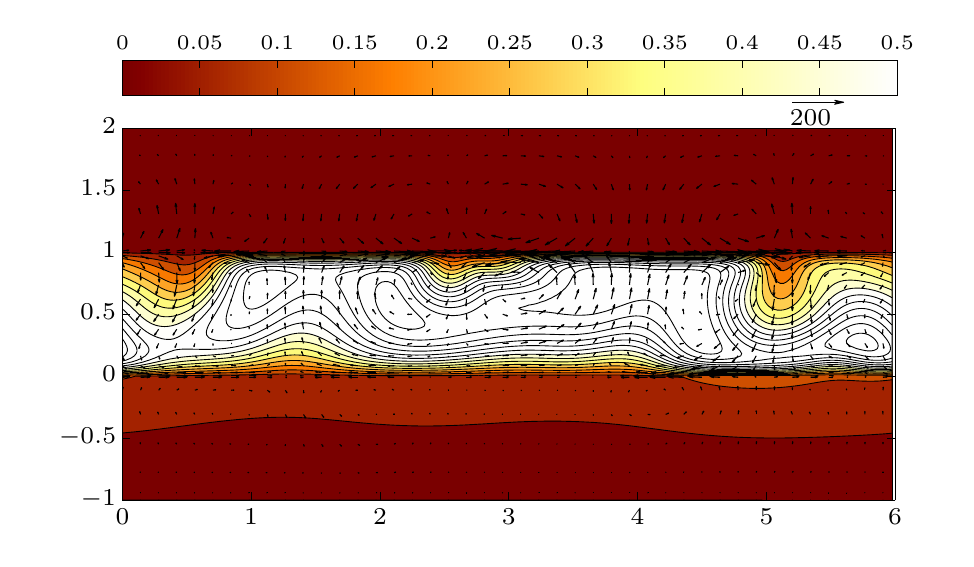} &
\includegraphics[ trim = 10px 10px 10px 10px, clip, width=0.35\textwidth]{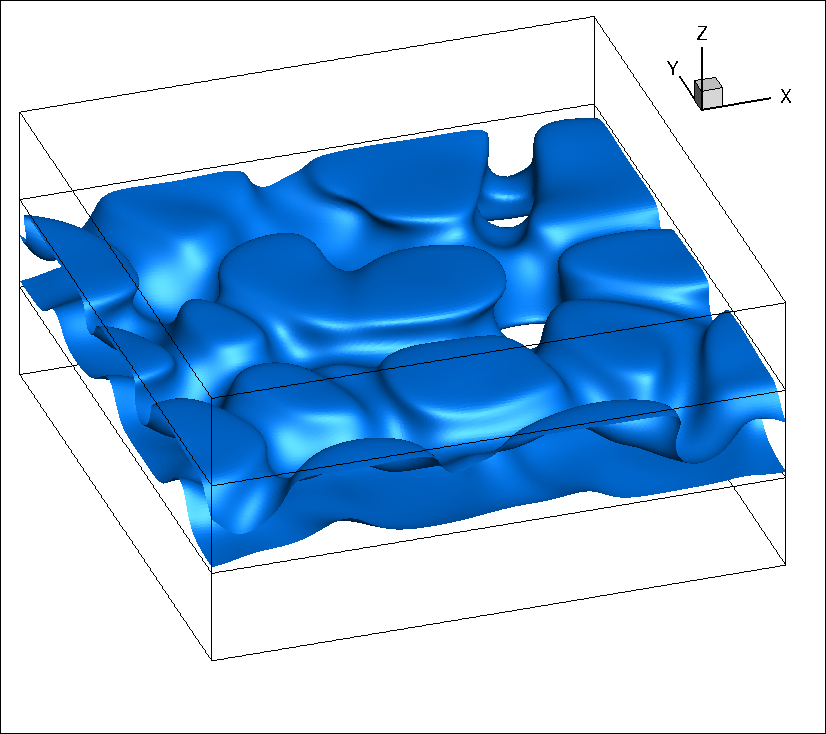}
\setlength{\unitlength}{0.5\textwidth}
\begin{picture}( 0, 0)(0,0)
\put(-1.93 , 1.37){\footnotesize (a)}
\put(-0.87 , 1.37){\footnotesize (b)}
\put(-1.93 , 0.55){\footnotesize (c)}
\put(-0.87 , 0.55){\footnotesize (d)}
\end{picture}
\end{tabular}
\caption{Marangoni convection $(G=0)$ in the reference configuration H4. See Fig.~\ref{fig:pattterns:h4noMa} for details. 
Here, the temperature isosurface in (d) is taken at $T=0.5$.}
\label{fig:pattterns:h4noG} 
\end{figure*}
%-----------------------------------------------------

The rms velocity in each of the three layers is plotted in  Fig.~\ref{fig:ref:integral} for RC (top panel), MC (middle panel), and RMC 
(bottom panel). After the onset of convection, all systems run into a statistically stationary state. For the
MC case the flow velocity is weakest. The differences in the rms  velocity between RMC and RC remain small, but become better  
visible when plotting the $x,y$-averaged vertical profiles, $\sqrt{\langle \mbf u ^2 \rangle_{x,y,\tau_{av}}}$, which is done 
in the top panel of Fig.~\ref{fig:ref:integral1}. The profiles are averaged over a time interval $\tau_{av}$, which is at least 10 convective 
time units long. 

For RC and RMC, the primary flow structure is caused by buoyant convection in the middle layer. Interestingly for the RMC case, the 
rms velocities are smaller in the outer layers than for RC. For MC, the flow is driven by interfacial tension gradients and therefore 
rms velocity amplitudes are largest at the interfaces. The lower interface contributes less because of the ratio $\xi=2.57=\Ma^{II}/\Ma^{I}$. For these 
three cases, the middle layer velocity has on average always the highest amplitude while the rms velocity in the bottom electrode is
always smallest. In cases RMC and RC,  the small rms velocity magnitude is most probably caused by the stable density stratification 
in the lower layer. In case of MC, the upper-layer and middle-layer rms velocity are of similar magnitude, which is also in line with the 
specific driving mechanism, namely that interface II between molten salt and upper electrode is the main source for Marangoni convection. 

The impact of convection on the $x,y$- and time-averaged, vertical temperature profiles is displayed in the bottom panel Fig.~\ref{fig:ref:integral1}. 
The temperature is smallest in the case of RMC and highest in the case of MC. In connection to this observation, we measure the global 
transport of heat by the Nusselt number in the following way. Following the works on convection with internal heating in a single 
layer by \citet{goluskin2016penetrative}, we define a Nusselt number Nu by the maximum of the $x,y$-averaged temperature,
$T_{max}$. It is given by  
%-----------------------------------------------------
\beq
 T_{max}(t) = \max \{ \langle  T \rangle_{xy}(z,t) : -1 \leq z \leq d_{21}+d_{31}\}.
\eeq
%-----------------------------------------------------
Furthermore, one takes the maximum temperature for pure conduction $T_{max}^{con}$ which is given by Eq.~\eqref{eq:tempmax:cond} 
and arrives at the definition 
%-----------------------------------------------------
\beq
{\Nu}= \frac{  T_{max}^{con}}{T_{max}}
\label{Nuss_internal}
\eeq 
%-----------------------------------------------------
of the Nusselt number, which will be used in the following.
Figure \ref{fig:ref:integral2} shows the Nusselt number as a function of time for the three convection cases. The difference between the three 
cases supports our observation for the temperature profiles in Fig.~\ref{fig:ref:integral1}. The initial drop is caused by the initialization with zero
temperature everywhere. The three-layer system requires a finite time to heat up. This heating time can be estimated 
from the balance between the time derivative of temperature and the heating rate  in Eq.~\eqref{eq:thermal:lay2:nodim}, which
provides a heating period of about $ \Pr^{(1)}{d_{21}}^2/(8 \kappa_{21}) \approx 0.1$ The reason of why RMC transports 
the heat more efficiently than RC can be answered by an examination of the typical flow structures, which is done next.

\begin{figure*}
\begin{tabular}{cc}
\includegraphics[ width=0.5\textwidth]{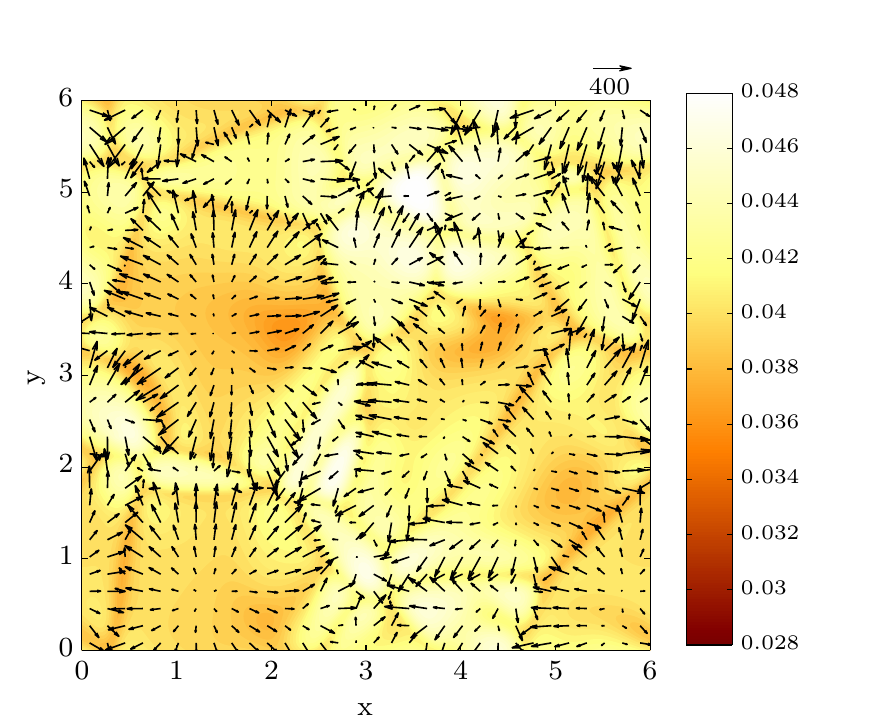} &
\includegraphics[ width=0.5\textwidth]{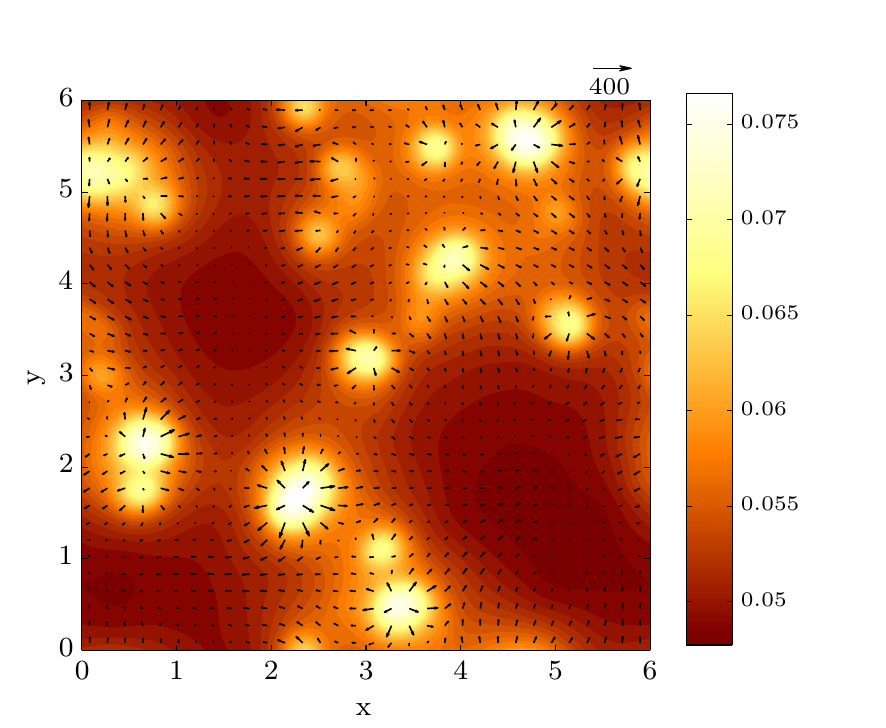} \\
\includegraphics[ width=0.55\textwidth]{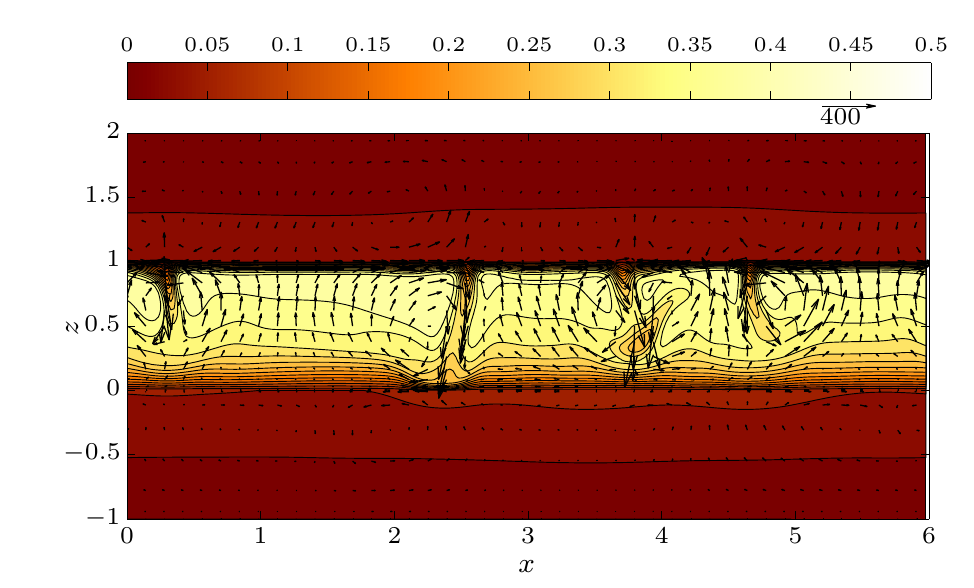} & 
\includegraphics[trim = 10px 10px 10px 10px, clip, width=0.35\textwidth]{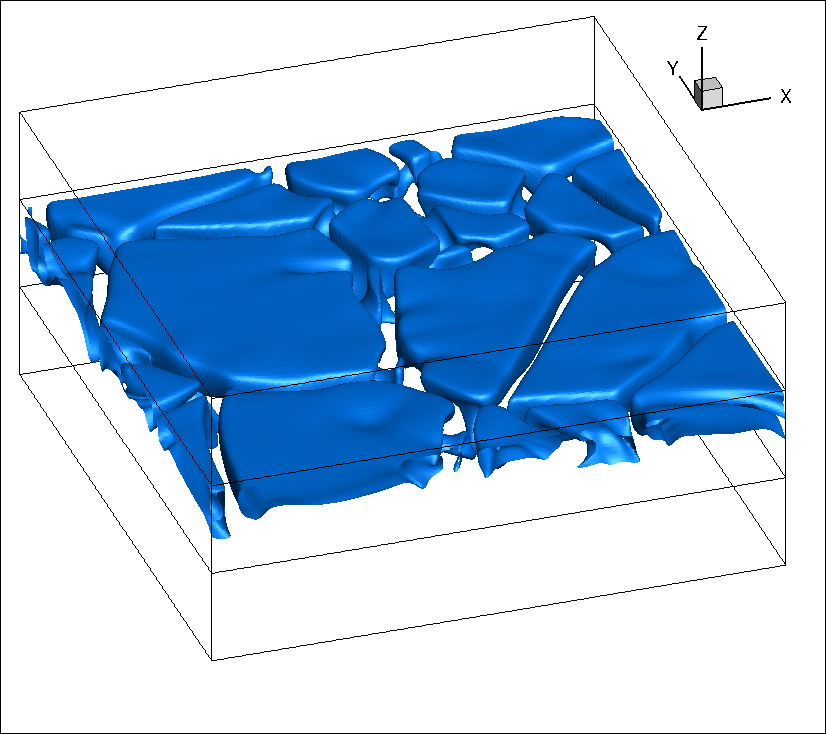}
\setlength{\unitlength}{0.5\textwidth}
\begin{picture}( 0, 0)(0,0)
\put(-1.93 , 1.37){\footnotesize (a)}
\put(-0.87 , 1.37){\footnotesize (b)}
\put(-1.93 , 0.55){\footnotesize (c)}
\put(-0.87 , 0.55){\footnotesize (d)}
\end{picture}
\end{tabular}
\caption{Rayleigh-Marangoni convection in the reference configuration H4. See Fig.~\ref{fig:pattterns:h4noMa} for details. Here,
the temperature isosurface in the lower right panel is taken at \tk{$T=0.37$ also}.}
\label{fig:pattterns:h4} 
\end{figure*}

In Figs.~\ref{fig:pattterns:h4noMa}, \ref{fig:pattterns:h4noG} and \ref{fig:pattterns:h4}, we display snapshots of the flow and temperature from 
different perspectives for cases RC, MC and RMC, respectively. All figures are taken at time $t=0.3$. In the contour plots at the upper interface II 
as well as in the vertical cuts at $y=0$, we kept the same temperature range in all three cases. The differences were too large for the lower
interface I such that we had to take different contour level intervals here. 

Case RC is organized in irregular cells with an upwelling hotter fluid in the center and a downwelling colder fluid at the cell boundaries, as seen
in all three two-dimensional cuts in Fig.~\ref{fig:pattterns:h4noMa}. The downwelling colder fluid appears in form of sheets and 
jets, which sometimes carry fluid all the way down to the lower interface I across the middle layer $\Omega^{(2)}$. 
In the vertical cut at $y=0$, one also observes that the upper and lower layers are coupled by the viscous stresses to the mid layer since the 
velocity fields are aligned. The isosurface of $T^{(2)}=0.37$ in the bottom right panel completes the description of the convection structures. Grooves 
are formed by downwelling cold fluid. They resemble the typical patterns which are visible in the temperature at the upper interface, $T^{II}(x,y,d_{21},t=0.3)$.

Case MC which is displayed in Fig.~\ref{fig:pattterns:h4noG} shows slightly different flow structures at the same time. The temperature distribution 
at the upper interface II depicts convection cells that are driven by the increase of interfacial tension from the cell center (hot) to the cell 
boundary (cold). The Marangoni effect also determines the flow behavior at the lower interface, i.e., the fluid streams from the hotter to colder 
regions. However, the structures across the middle layer are strongly influenced by the convection cell patterns that are formed at the upper interface. 
The isosurface of $T=0.5$ in the bottom right panel Fig.~\ref{fig:pattterns:h4noG} encloses hotter fluid. It is seen how fluid is transported from the 
center of the middle layer to the upper interface where it cools down. The vertical cut of the figure complements the view on the convection structure
in this case.

Case RMC resembles mostly case RC, which can be seen by the similar composition of temperature distributions in the top left panel of
Fig.~\ref{fig:pattterns:h4}. However, there are some slight differences visible. The convection cells in the top left panel have a smaller typical 
length scale. This is the reason for the enhanced heat transport and the higher Nu which was shown in Fig.~\ref{fig:ref:integral2}.  Moreover, downwelling fluid appears preferentially in form of stronger jets rather than 
sheets. These structures couple the top and bottom electrode directly. The more frequent jets are also visible by the hot spots in the 
lower interface in the top right panel and in the isosurface in the lower right panel. This is because cold regions -- the source of downwelling jets 
-- have a tendency now to merge by the Marangoni effect. In conclusion, the flow pattern of RMC is of smaller length scale than the one for pure RC. This decrease 
of length scales is to our view the reason for the enhanced Nusselt number since the heat can be carried along a larger number of up- and downwelling 
structures. 
%-----------------------------------------------------
%-----------------------------------------------------

\subsection{Variation of the layer height in H1 to H6}
After the detailed description of the reference case H4, we turn in the following to the variation of the layer heights.
In the simulation series H the ratios $d_{21}$ and $d_{31}$ will be kept to unity and the electrical current density remains
fixed to 3kA/$m^2$ (see cases H1 - H6 in Tab.~\ref{tab:paramteric-study:height}). The physical height $d^{(1)}$ will be 
changed.  For case H1 with $d^{(1)}=5$ mm layer, the system is linearly stable and all initial infinitesimal perturbations 
decay with respect to time. As already discussed in Sec.~\ref{sec:res:ref}, the simulations run always at least 10 convective
time units for any statistical analysis. In addition to each of the RMC cases, which are given in Tab.~\ref{tab:paramteric-study:height}, 
corresponding RC runs are also conducted.
%-----------------------------------------------------
\begin{figure}
\includegraphics[ width=0.45\textwidth]{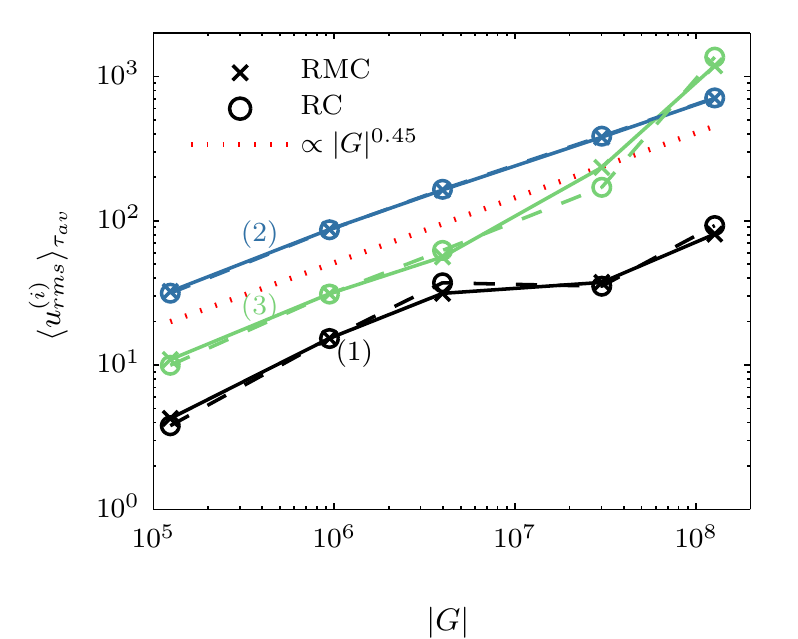} 
\caption{Root mean square velocities as a function of the magnitude of the Grashof number, which is directly connected
to the variation of the reference height $d^{(1)}$. Solid lines are for RMC and dashed lines for RC. Each of the three layers
is indicated in the plot.}
\label{fig:ma:var:height} 
\end{figure}
%-----------------------------------------------------
\begin{figure}
\includegraphics[ width=0.4\textwidth]{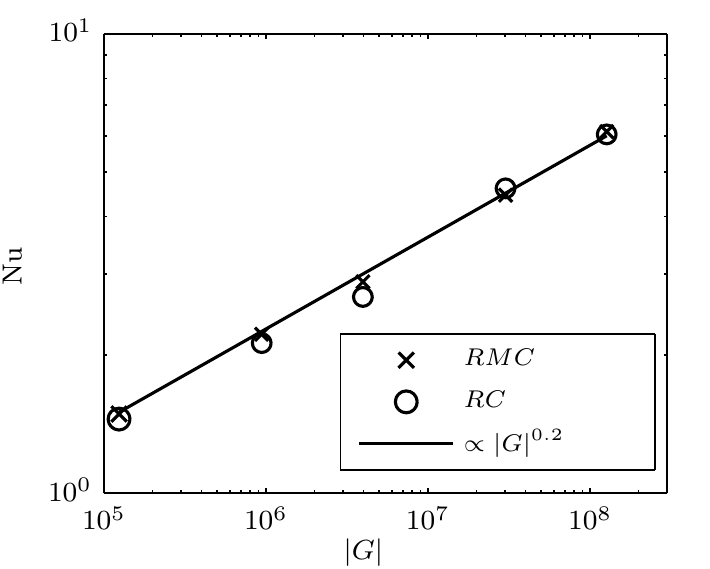}
\caption{Global transport of heat (bottom panel) versus Grashof number magnitude for DNS series 
H1 to H6 as measured by the Nusselt number for the whole system (\ref{Nuss_internal}).}
\label{fig:ma:var:height1} 
\end{figure}
%-----------------------------------------------------

Fig.~\ref{fig:ma:var:height} displays the rms velocity in each of the three layers. As already described for the reference case, 
the strongest flow is observed in the middle layer, except in case H6. For the tallest system, the upper layer shows the highest 
magnitude of rms velocity. The reason lies in the unstable stratification in the upper layer. It  is strong enough to develop an 
additional large-scale convection cell (in the sense of Rayleigh-B\'{e}nard convection) that contributes dominantly to the rms 
magnitude. This behavior can also be rationalized by the Rayleigh number $\Ra^{(3)}$ for the upper layer $\Omega^{(3)}$ which 
takes values of $-193$ for H4, $-1469$ for H5 and $-6192$ for case H6, respectively.  As visible from Eq. \eqref{eq:upperlayer:Rayleigh}, 
case H6 is far beyond any linear instability threshold. Thus, a significant convective fluid motion can be
expected to exist for this case only. 

We also indicate a power law scaling of $G^{0.45}$ in Fig.~\ref{fig:ma:var:height} that matches the data for the rms velocities in 
the molten salt layer very well. All computed values are also listed in Tab.~\ref{tab:results:dns} in the appendix. The Reynolds number is 
a measure for the global transport of momentum in turbulent convection flows. The definition for each of the three layers in the present 
liquid metal battery model is given by 
\begin{align}
\Rey^{(1)}&= \langle u_{rms}^{(1)} \rangle_{\tau_{av}}\,,
\label{GlobalMom1}\\ 
\Rey^{(2)}&= \frac{\langle u_{rms}^{(2)}\rangle_{\tau_{av}} d_{21}}{\nu_{21}}\,,
\label{GlobalMom2}\\ 
\Rey^{(3)}&= \frac{\langle u_{rms}^{(3)}\rangle_{\tau_{av}} d_{31}}{\nu_{31}}\,.
\label{GlobalMom3}
\end{align} 
Thus, the Reynolds numbers will exhibit exactly the same trend as the rms velocities in Fig.~\ref{fig:ma:var:height} and are therefore not shown.
Only the relative amplitudes with respect to each other vary since the liquid metal has lower kinematic viscosity than the molten salt.
In classical turbulent Rayleigh-B\'{e}nard convection with no-slip and isothermal walls for air \cite{Schumacher2014}
or liquid metals \cite{Scheel2016} as working fluids the Reynolds numbers obeys nearly the same scaling with increasing Grashof 
number. The scaling deviates slightly from the exponent of 1/2. The trends for the Reynolds numbers in both liquid metal electrodes 
vary stronger and cannot be fitted by a power law.  The reason for this behavior can  most probably be attributed to the more complex 
boundary conditions at the interfaces which couple the motion between separator and electrode.

\begin{figure}
\includegraphics[ width=0.48\textwidth]{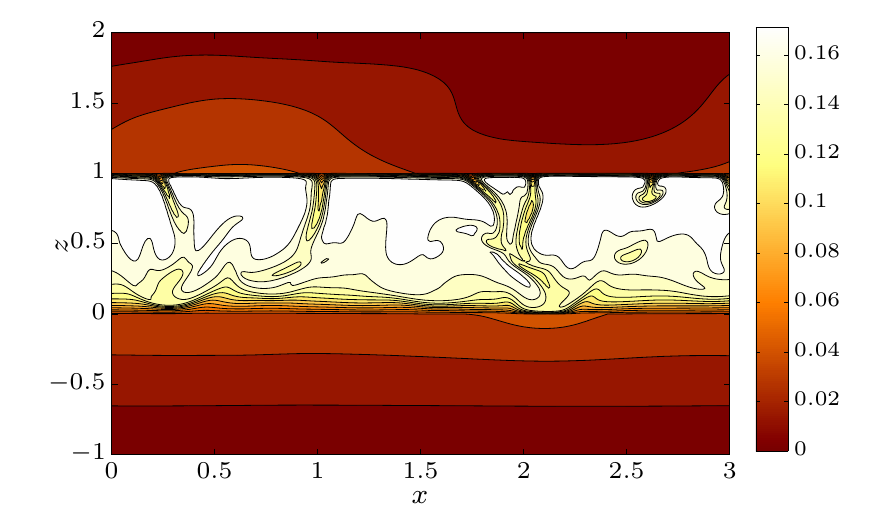}
\caption{Vertical instantaneous cut through the three-layer battery model for case H6 at \tk{ $y=0$, $t=0.04$. 
Temperature field contours are plotted.}}
\label{fig:ma:var:height1a} 
\end{figure}

The scaling of the Nusselt number -- the measure for the global transport of heat -- with respect to the Grashof number is displayed 
in Fig.~\ref{fig:ma:var:height1}. A power law 
exponent of $0.2$ as suggested for Rayleigh-B\'{e}nard convection with internal heat sources by \citet{goluskin2016penetrative} fits
well to the present simulation data. RMC has generally a slightly higher Nusselt number in comparison to RC, which is a result of the additional
Marangoni effects. The values for the Nusselt numbers are also listed in Tab.~\ref{tab:results:dns} in the appendix. Note also that the 
Marangoni number varies as
$\Ma\propto G^{3/5}$ for the present cases of variable layer height. Figure \ref{fig:ma:var:height1a}
shows a vertical cut through the three-layer system of case H6. The temperature contours indicate a strong temperature gradient at the 
interface II. It can also be seen that the thermal boundary layers at both interfaces have different thicknesses. 

\begin{figure*}
\begin{tabular}{cc}
\includegraphics[ width=0.5\textwidth]{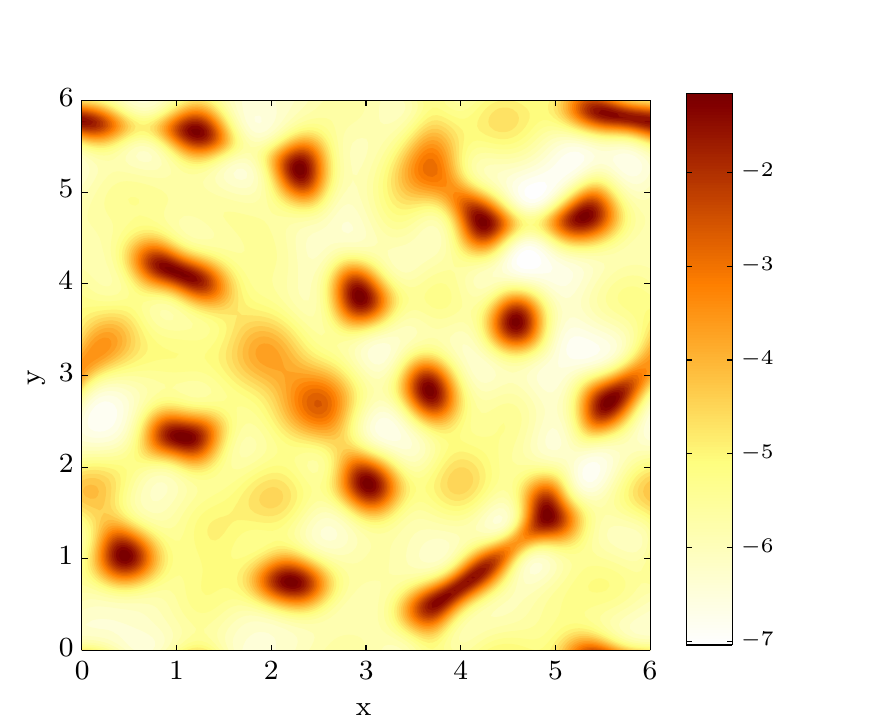} &
\includegraphics[ width=0.5\textwidth]{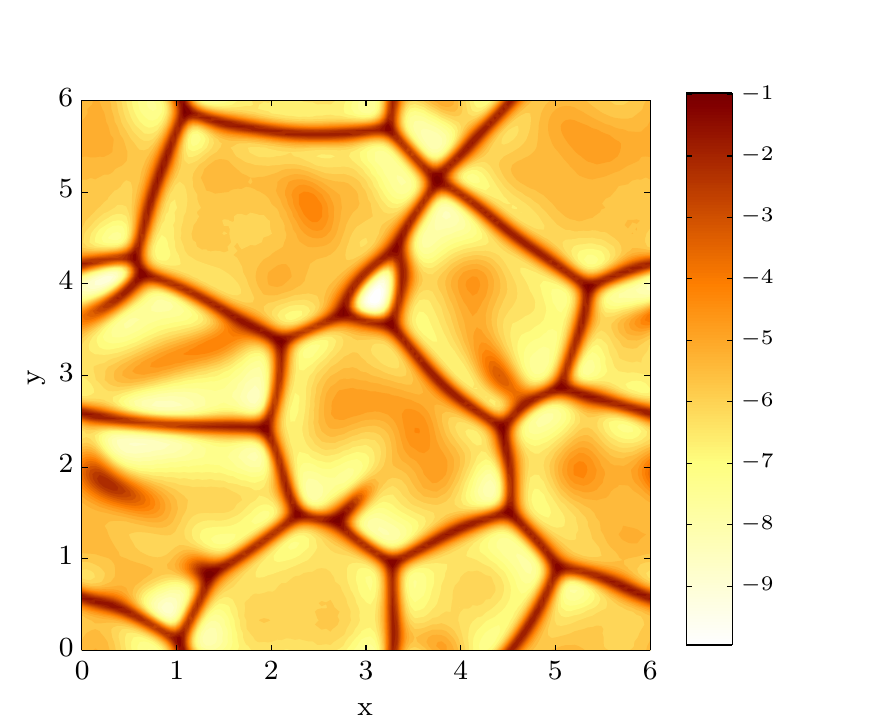}
 \\
\includegraphics[width=0.5\textwidth]{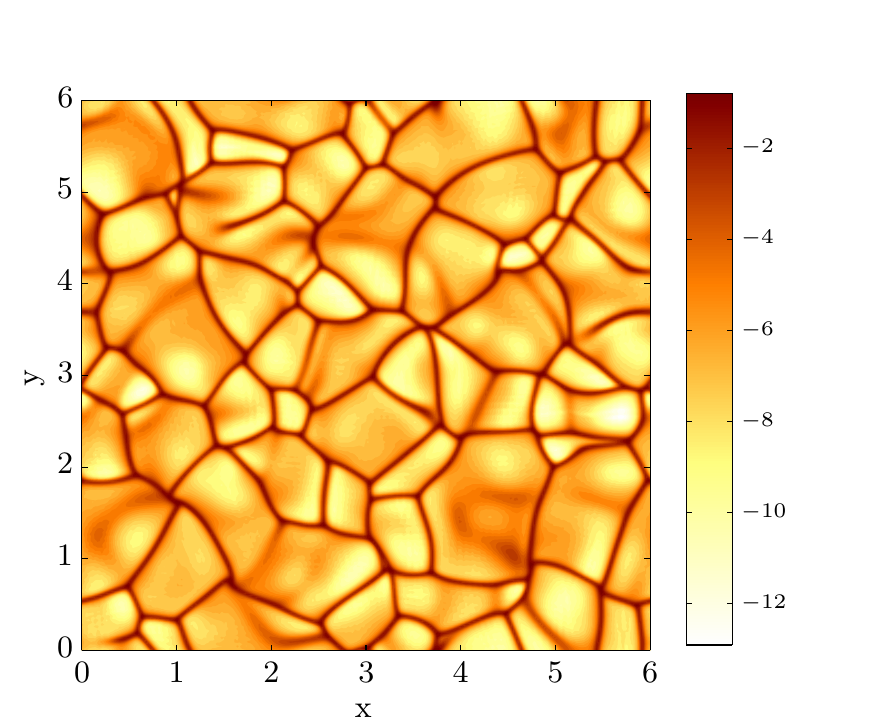} & %%% this is still to be updated
\includegraphics[ width=0.5\textwidth]{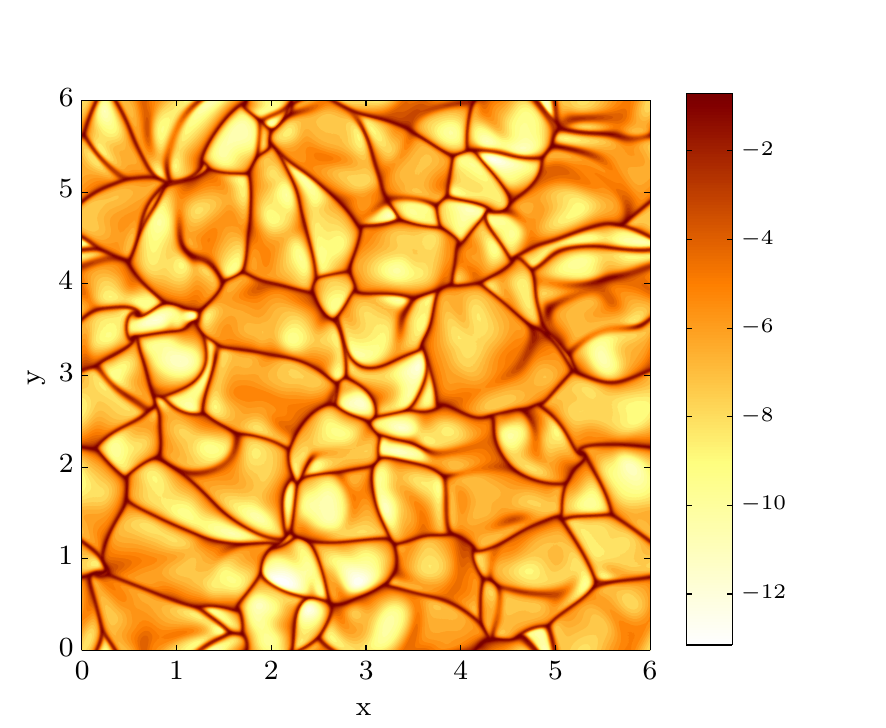}
\setlength{\unitlength}{0.5\textwidth}
\begin{picture}( 0, 0)(0,0)
\put(-2.05 , 1.55){\footnotesize (a)}
\put(-1.03 , 1.55){\footnotesize (b)}
\put(-2.05 , 0.7){\footnotesize (c)}
\put(-1.03 , 0.7){\footnotesize (d)}
\end{picture}
\end{tabular}
\caption{Vertical temperature gradient at the upper interface I. Contours of $\partial_z T^{(2)}(x,y,z=d_{21})$ for simulation 
runs with variable middle layer height are shown: Top left(a) case H2 at t=0.5; Top right(b) case H3 at $t=0.3$; Bottom left(c) 
case H5 at $t=0.2$; Bottom right (d) case H6 at $t=0.04$ }
\label{fig:var:height:2} 
\end{figure*}

The variation of the convection pattern with height is represented by the vertical gradient of the temperature  
$\partial_z T^{(2)}(z=d_{21})$ at the upper interface as seen in Fig.~\ref{fig:var:height:2}. We use this quantity because it displays
a larger contrast between inflow and outflow regions in comparison to a plot of the interfacial temperature. At the bright temperature 
contour areas one observes inflow and at the darker temperature contour areas outflow of the fluid from the interface. With increasing 
Grashof number, the cellular pattern is fragmented into ever finer substructures such that a larger amount of heat can be carried 
across the layer. 

%-----------------------------------------------------
%-----------------------------------------------------
%-----------------------------------------------------

\subsection{Joint variation of the middle-layer size and the electrical current density}
In the present subsection, we will vary the relative thickness of the middle layer, $d_{21}$. It will be reduced to 10mm, which corresponds 
to $d_{21}=0.5$ in series JM, to 6mm ($d_{21}=0.3$) in series JS, and to 2mm ($d_{21}=0.1$) for series JXS. 
The outer electrode heights will remain unchanged at a value of 20mm. Moreover, for each of these three geometries, different 
current densities $j_0$ were taken as summarized in Tab.~\ref{tab:paramteric-study:height}. 

Let us start with the medium sized middle layer at $d_{21}=0.5$. All four cases JM1 - JM4  are unstable to Rayleigh-B\'{e}nard 
convection which is readily predicted by a comparison of $G$ with the corresponding critical Grashof numbers $\tilde G_c$ in 
Tab.~\ref{tab:paramteric-study:height}. Again, we simulated the cases of RMC as listed in the table 
together with corresponding RC simulations with $Ma=0$ in order to emphasize the effects of the interfacial tension better.

With a further reduction of the thickness of the middle layer $\Omega^{(2)}$ the corresponding Rayleigh-B\'{e}nard and Marangoni 
modes are damped to the subcritical range as they rely on convection in this layer. Thus a higher current density of 
$j_0$=5kA/$m^2$ and consequently a higher heating rate are required to sustain convection (see also Tab.~\ref{tab:paramteric-study:height}).
The case JXS1 with the shallowest middle layer remains stable although the current density is increased to a value as high as $j_0=20$kA/m$^2$.
Only when the outer layer sizes are increased to $d^{(1)}=40$mm, the system becomes unstable again and shows the upper layer Rayleigh mode
(case JXS2). 

\begin{figure}
\includegraphics[ width=0.4\textwidth]{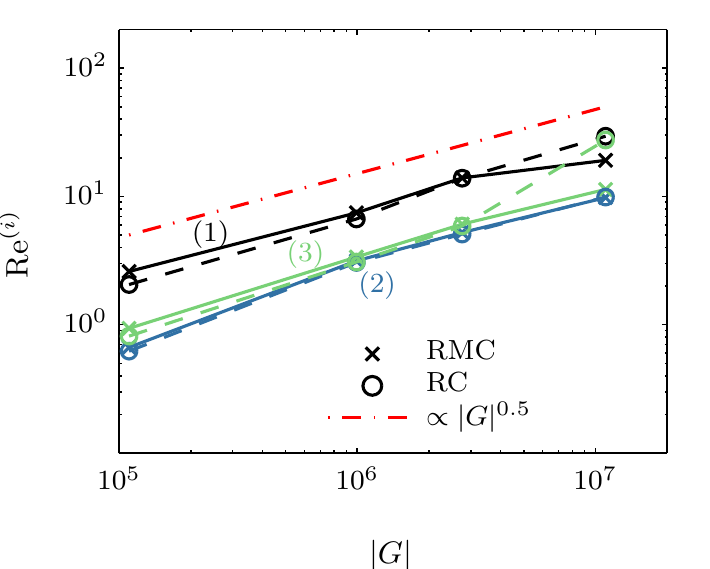}\\
\includegraphics[width=0.4\textwidth]{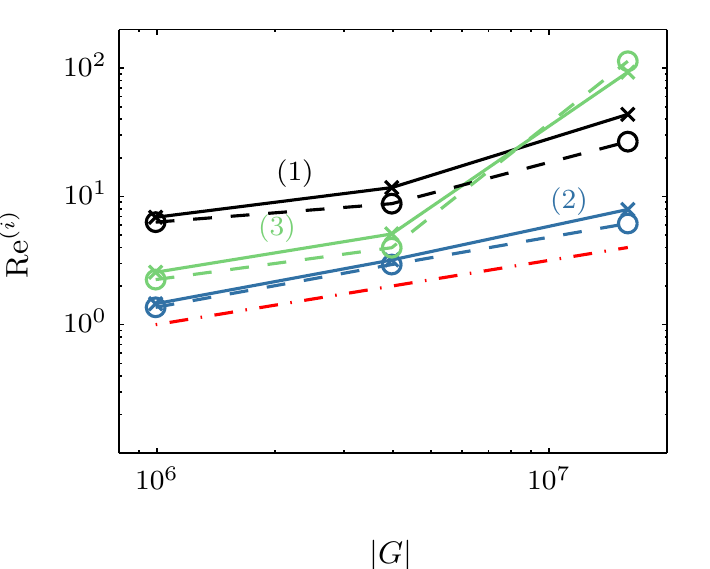}
\setlength{\unitlength}{0.5\textwidth}
\begin{picture}( 0, 0)(0,0)
\put(-0.85, 0.6){\footnotesize (b)}
\put(-0.85, 1.25){\footnotesize (a)}
\end{picture}
\caption{Reynolds number scaling versus Grashof number for shallower middle layers and different electrical current densities with
RMC (solid line) and RC (dashed line). Top panel (a): Reynolds numbers $\Rey^{(i)}$ for series JM, Bottom panel (b): Reynolds number for series 
JS. As a guide to the eye, we also plot $\Rey\sim |G|^{0.5}$ as a dash-dotted line.}
\label{fig:ma:var:current} 
\end{figure}

The observations from these three additional series of simulation runs are as follows. 
Figure~\ref{fig:ma:var:current} (top panel for JM and bottom panel for JS) shows the Reynolds numbers $\Rey^{(i)}$ as a function of the Grashof 
number magnitude $|G|$. The data for both series follow again a similar power law trend as in the previous parameter study for H1 to H6 with an exponent 
that is close to 1/2. The Reynolds numbers $\Rey^{(3)} $ for cases with the largest Grashof numbers (JM4 and JS5)  show a significant
enhancement. This large amplitude of the Reynolds number is caused by the strong Rayleigh-B\'enard mode which is established in the 
upper liquid metal electrode. The corresponding Rayleigh number $\Ra^{(3)}$, which is defined in Eq.~\eqref{eq:upperlayer:Rayleigh},
is $\Ra^{(3)}=-1092$ for JM4 and $\Ra^{(3)}=-2673$ for JS5. 

\begin{figure}
 \includegraphics[ width=0.4\textwidth]{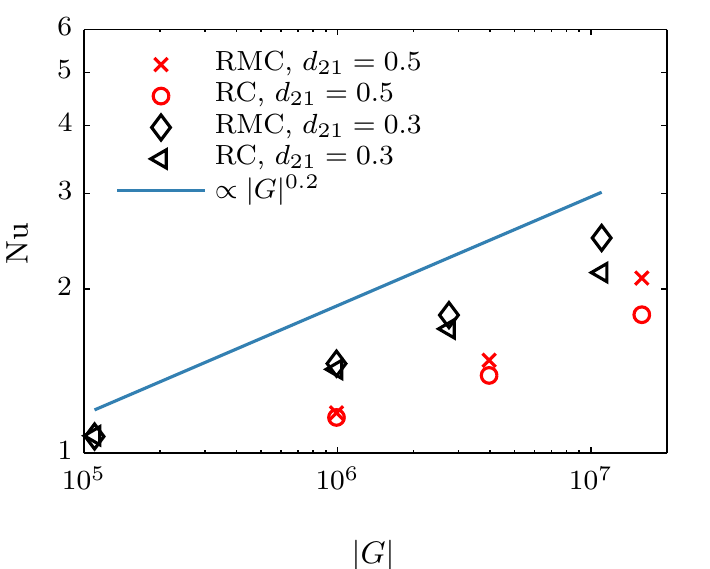} \\
 \includegraphics[ width=0.4\textwidth]{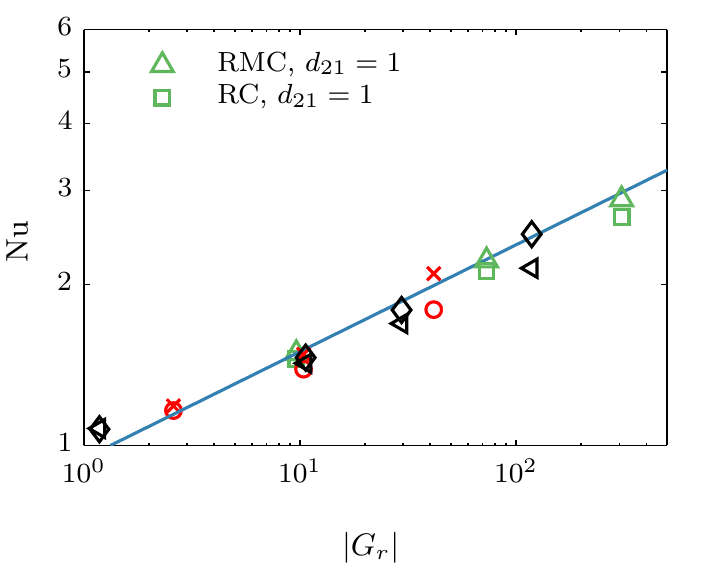}
\setlength{\unitlength}{0.5\textwidth}
\begin{picture}( 0, 0)(0,0)
\put(-0.89, 0.6){\footnotesize (b)}
\put(-0.89, 1.25){\footnotesize (a)}
\end{picture}
\caption{Nusselt number scaling versus Grashof number for shallower middle layers and different electrical current densities. 
Top panel (a): Nusselt numbers for JM and JS series. Bottom panel (b): all Nusselt numbers replotted as a function of the relative 
Grashof number $G_r=G/\tilde G_c$.}
\label{fig:ma:var:current1} 
\end{figure}
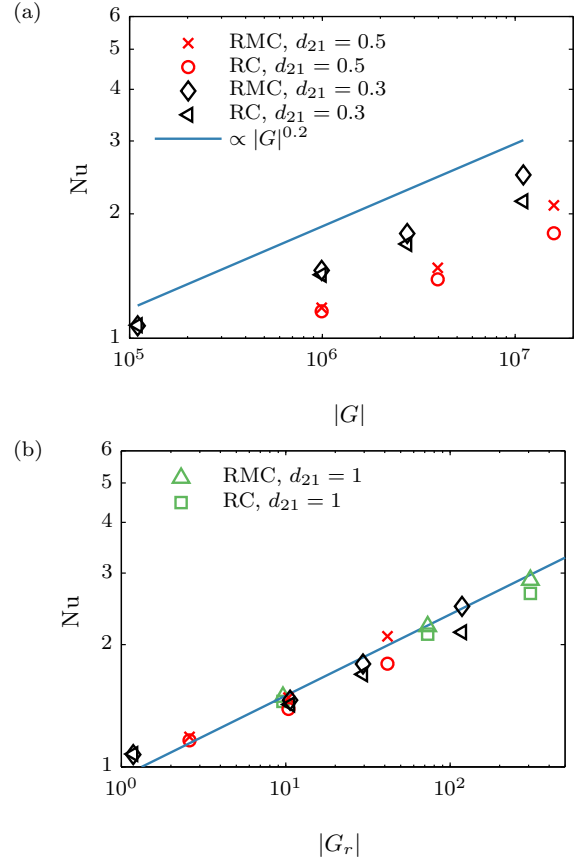

Figure \ref{fig:ma:var:current1} shows the Nusselt numbers for simulation series JM and JS. It is seen that $\Nu$ is systematically larger for the 
shallower middle layer at $d_{21}=0.3$ when plotted against the Grashof number. Also, RMC has a consistently higher value of $\Nu$ than 
RC. An approximately universal scaling relation for the turbulent heat transfer, for which all data points collapse to a single line in a double-log plot, 
can be obtained when the original Grashof number is substituted by the \textit{relative Grashof number} $G_r$. It is defined as the actual 
Grashof number divided by the corresponding critical value $G_r= G/ \tilde G_c$. This relation is shown in the bottom panel of 
Fig.~\ref{fig:ma:var:current1} for all simulation cases in series JM and JS. The data are fitted well again by the power law of $\Nu\propto G^{0.2}$ that was 
discussed already in section IV B. This particular scaling of the turbulent heat transfer seems thus to be very robust in our parametric studies. 

%--------------------------------------------------------
%--------------------------------------------------------
%--------------------------------------------------------

\section{Different convection regimes}
\label{sec:diss}
As the full nonlinear simulations  in the present three-layer model showed, there are four different modes  for thermal convection that we have already
discussed, and an additional one  not discussed so far. These are (1) the Rayleigh-B\'{e}nard mode due to internal heating in the middle 
layer, (2) the standard Rayleigh-B\'{e}nard mode in the upper layer due to the hot interface II and the cold top, (3) the Marangoni mode 
at the upper interface II, (4) the Marangoni mode at the lower interface I, and (5) an anti-convection mode in the middle layer and bottom electrode  
that has not been discussed so far. On the one hand, these modes are partly coupled to each other, one the other hand, they are partly insignificant
for particular parameter sets. They contribute differently to the global transfer of heat and momentum. 
The following section will discuss and summarize these different mechanisms and how their appearance can be predicted.

\subsection{Rayleigh-B\'{e}nard mode in molten salt layer}
The primary mechanism for convection in the investigated system is the unstable temperature distribution between the maximum 
temperature close to $z=d_{21}/2$  and the interfacial temperature at the upper interface II, causing buoyant convection. In a 
classical study, \citet{sparrow1964thermal} investigated the stability of a single fluid layer under internal heating, with no-slip and 
isothermal boundary conditions at the top and bottom. These results of \citet{sparrow1964thermal} for one layer can be translated 
to the present three-layer system. Aside from the formal linear stability of a layer with no-slip and isothermal boundaries, they found 
that the threshold of convection can be described by a single Rayleigh number criterion if the interface temperatures do not differ too 
much. The relevant Rayleigh number $\Ra^{(2)}$  relates the difference of the temperature maximum and the upper interfacial temperature 
and their distance. Formally one can thus define the Rayleigh number as follows
\beq
\Ra^{(2)}=\dfrac{g \Theta(T_{cond}^{max}-T^{II}_{cond}) \beta_T^{(2)} \left(d^{(2)}- \dfrac{d^{(2)}B}{2A}\right)^3 }{\nu^{(2)} \kappa^{(2)} }, 
\label{eq:rayleigh:mid:1}
\eeq
where the distance of the temperature maximum to the interface is expressed with the help of \eqref{eq:pos_max_temp}. 
In terms of this Rayleigh number \eqref{eq:rayleigh:mid:1}, \citet{sparrow1964thermal} showed that the threshold for the linear 
instability falls in a range of $\Ra^{(2)}= 560$ -- 595 (depending on the outer temperature difference which they denoted by a
parameter $N_s$). Later, \citet{char1994stability} studied the linear stability of a single layer subject to internal heating with different 
boundary conditions for the linearized Boussinesq equations. The authors derived a critical Rayleigh number for a free-slip, isothermal and free-slip, 
thermally insulated upper boundary, respectively. They derived a critical Rayleigh number of $\Ra^{(2)}$=296.20 (112.64) for the isothermal 
(thermally insulated) case. Which boundary conditions are better suited to account for layer (3) is unclear. A hint for the best thermal 
boundary condition might be given by the ratio of thermal conductivities $\lambda_{32}=137.54$.
This ratio suggests an isothermal rather than an thermally insulating boundary condition.

To check the relevance of this newly defined Rayleigh number $\Ra^{(2)}$, the calculated critical Grashof numbers can be transformed 
to critical Rayleigh number for a fixed set of material parameters and geometry. For the four different geometries which are given in 
Tab.~\ref{tab:paramteric-study:height}, one obtains a critical Rayleigh number of $\tilde \Ra^{(2)}_c= -290.7$ for $d_{21}=1$,  
$\tilde \Ra^{(2)}_c= -281.7$ for $d_{21}=0.5$,  $\tilde \Ra^{(2)}_c=-276.9$ for $d_{21}=0.3$ and $\tilde \Ra^{(2)}_c=-92.7$ for $d_{21}=0.1$. 
Obviously, for the thick middle layer the critical Rayleigh number 
fits thus well to the prediction of \citet{char1994stability} of an isothermal upper interface. When the middle layer thickness is reduced, the upper-layer 
Rayleigh-B\'{e}nard mode gets more involved and reduces the stability threshold. For a shallow middle layer, the internal convection mode gets 
insignificant. At this point we recall that the separator layer should be thin to keep the Ohmic losses at a minimum   

Finally, let us note that in the present case, the number of parameters in the marginal stability analysis, which can be varied independently 
of each other in the nondimensional equations, is less than 18. The stability problem in Sec.~\ref{sec:stab:problem} 
can be rescaled to a new velocity $w^+=w \Pr$. By doing this, one observes that only the product $\Ra=G\,\Pr$ appears in the linearized 
equations for perturbations with a growth rate of zero. Furthermore, the conductivities of the outer layer can be regarded as infinite as 
is done in the solution for pure conduction. Overall, this results in a reduction to 15 essential parameters that govern linear stability 
threshold.

\subsection{Rayleigh-B\'enard mode in upper liquid metal electrode}
The top electrode has a linear temperature profile, and thus its stability to buoyant convection can be estimated by the Rayleigh number 
$\Ra^{(3)}$ which was  introduced in Eq.~\eqref{eq:upperlayer:Rayleigh}. For the simulated geometries, this Rayleigh-B\'{e}nard mode was 
the primary source for convection in the series of the very shallow cases, JXS. According to our stability calculation the critical Rayleigh 
number for $d_{21}=0.1$ is $\Ra^{(3)}_c=-1290.4$. For thicker middle layers the convection mode is observed for higher electrical current
densities only.  The Rayleigh number for which convection is observed is $\Ra^{(3)}=-2673$ for JS5,  $-1092$ for JM4 and 
 $-6192$ for H6. In summary, if shallow middle layers are combined with thicker layers for the upper electrode, this specific Rayleigh-B\'{e}nard 
mode can become the primary source for convection in the three-layer system. If the internal convection mode in the molten salt layer is 
active additionally, this Rayleigh-B\'{e}nard mode leads to a considerable increase of the velocity fluctuations in the top layer. This parameter regime
would be an example where both RC flows dominate the global transport. 

\subsection{Marangoni modes at the liquid metal-molten salt interface}
The general requirement for the existence of stationary MC can be inferred from the stability theory of two layers by \citet{sternling1959interfacial}. 
The respective interface is prone to MC if heat is transferred from the phase with lower thermal diffusivity into the phase with higher one, given  
that interfacial tension decreases with temperature. Formally, the requirement for the respective interfaces in the present system is that 
$\kappa^{(2)}< \kappa^{(1)}$ and $\kappa^{(2)}< \kappa^{(3)}$. For the present liquid metal battery model both inequalities hold. Also, we can
expect that this ratio of diffusivities is typical for the compound system of a liquid metal and a molten salt in general since the diffusivity of the molten 
salt is by two orders of magnitude lower than that of the liquid metals in the electrodes.

In some of our cases, linearly unstable MC modes could grow, but the Rayleigh-B\'{e}nard mode due to internal heat sources was usually amplified 
stronger. We demonstrated that the main reason of the appearance of MC, as a part of the RMC regime, is a reduction of the length scales of the
cellular flow patterns which is in line with an increase of the Nusselt number.

The stability threshold of MC depends crucially on the ratio of transport coefficients of the adjacent phases. Thus, a threshold based on simple 
parameters, as in the case of the RC modes in the middle and upper layers, respectively, cannot be provided. It might be possible that MC is the 
primary mechanism for other sets of material parameters. This is stated in view of the missing experimental data for the interfacial tension 
of liquid metal--molten salt interfaces as a function of the temperature. What can be certainly stated here is that MC is damped by a reduction 
of the middle layer thickness (cf. Fig.~\ref{fig:ma:stab:3a}), by a reduction of the temperature difference and by an increased viscous friction, respectively. 
Thus a relevant Marangoni number should remain proportional to $(d^{(2)})^3$ as suggested by Fig.~\ref{fig:ma:stab:3a}(a).
The scaling with the third power of thickness is due to our choice of temperature scale.
Finally, we note that a closed expression for the stability threshold of pure MC with the generalized temperature profile of 
Eqns.~(\ref{eq:conduction:internalheating:1})--(\ref{eq:conduction:internalheating:3}) might be derived similar to the case of linear 
temperature profiles \citep{georis1993thermocapillary}.

\subsection{Anti-convection mode in the lower liquid metal alloy electrode}
The fifth source of convection concerns the bottom and the middle layer and was not discussed so far. Owing to the low Prandtl number 
$\Pr^{(1)}$=0.012 in the bottom layer, the diffusive transport of the temperature field dominates in this layer over advection for almost all 
simulated cases, which is readily estimated by the P\'{e}clet number $\Pe=\Rey^{(1)}\,\Pr^{(1)}<1$. This circumstance and the rather low 
heat capacity of the Pb-Bi layer will generate a diffusive transport into the bottom layer that is initiated at the localized hotspots due RC 
mode in the middle layer. This can be seen in Fig.~\ref{fig:ma:var:height1a} around $x=2$. The resulting horizontal temperature gradients 
cause a flow where denser colder fluid tends to displace hotter fluid. And indeed, the work that is done by gravity on the lower layer and 
which is proportional to $\langle -\beta_T^{(1)}T^{(1)}u_z^{(1)} \rangle_{xyz}$ is found to be positive for all simulated cases. 

Anti-convection has been classically described for two layers heated from above \citep{welander1964convective,boeck2008three,merkt2012pattern}. 
In this regime, which is noted as "stably stratified", buoyant convection is nevertheless \textit{possible} when the fluid mass density decreases 
with temperature, as it is usually  the case. The requirement for the anti-convection regime is that material parameters of both layers are in a 
certain proportion to each other as discussed by \citet{welander1964convective}. His prediction for a system with two layers heated from above 
can be transferred to the lower and middle layers in our case. We found that anti-convection would be possible if the thermal expansion coefficient 
of the middle layer is smaller than the one used for simulations. For a value of $\beta_{T, 21}=0.25$ instead of 2.59 anti-convection should become 
possible.   

\begin{figure}
\includegraphics[width=0.5\textwidth]{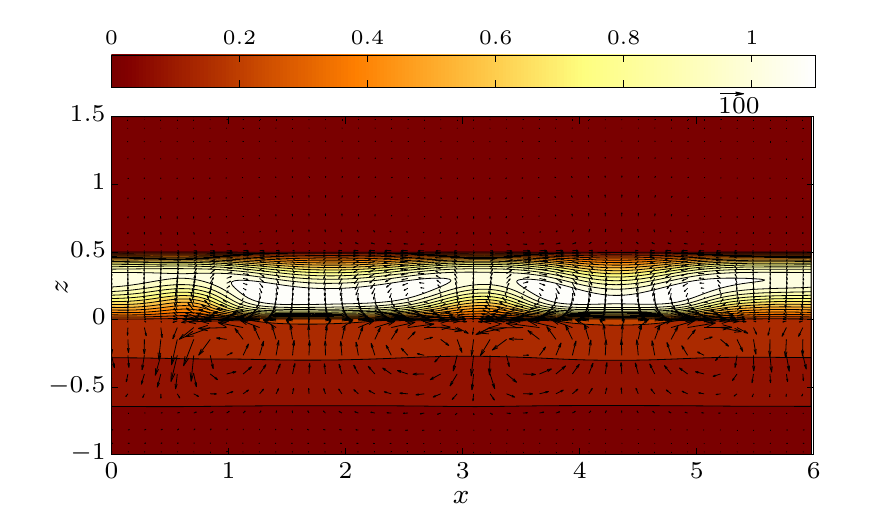} 
\caption{Anti-convection observed in a simulation for parameters of JM4, but with $\Ma=0$ and $\beta_{21}=\beta_{31}=0$. 
A vertical cut $y=0$ with velocity vectors and temperature contours at a time of $t=0.145$ is shown. This is shortly after the 
onset of anti-convection.}
\label{fig:anticonvection} 
\end{figure}

To demonstrate that this mode can indeed play a role in our case of internal heating in the middle layer, we performed a simulation with the parameters 
of run JM4 but suppress all sources for convection, i.e., we set  $\Ma=0$ as well as  $\beta_{21}=\beta_{31}=0$. The only physical mechanism which 
is kept is the change of fluid density and the resulting buoyancy effect in the lower layer. Figure~\ref{fig:anticonvection} demonstrates that indeed -- 
though the lower layer is stably stratified -- convective fluid motion sets in. In the middle layer, temperature perturbations are generated which 
heat the interface and produce an upwelling flow in the bottom layer. However, in the real system, the buoyancy in the middle layer counteracts 
while the Marangoni effect concurs with anti-convection since the interfacial flow is from the hot to cold regions. For the chosen parameters,
the present system develops a state of chaotic pulsating and translating convection cells. The time-averaged flow amplitude is considerable. One 
gets values of $u_{rms}^{(1)}=56.8$,  $u_{rms}^{(2)}=36.97$ and $u_{rms}^{(3)}=9.86$, respectively. 
%-------------------------------------------------
%-------------------------------------------------

\section{ Interfacial deformation }
The present model neglects effects of the deformation of the interfaces. We, therefore, need to verify
the consistency of if this assumption. To this end, we estimate the flow-induced deformation of the interface 
from the equilibrium shape. 

The shape of the interface is essentially governed 
by the balance of normal stresses at the interface. This condition was approximated by assuming zero vertical velocity.  
For a material interface, the relation by \citet{edwards1991interfacial} holds at any point on the interface. It is given by 
\beq
2H \sigma^{II} = \tilde p^{(2)}- \tilde p^{(3)}+ \mbf n \cdot \left( 2\mu^{(3)}  \hat{\mbf E}^{(3)}  - 2\mu^{(2)} \hat{\mbf E}^{(2)} \right) \cdot \mbf n\,, 
\label{eq:normal:stress}
\eeq
where all quantities carry a physical dimension here. Three new quantities have to be defined. First, the interface normal 
$\mbf n$ is pointing from phase (2) to phase (3) which is required by sign convention. This surface normal can be 
determined from a height function $h(x,y,t)$ that represents the distance of the interface from the unperturbed level, $z=0$, 
namely
\beq
\mbf n= \frac{-\partial_x h \mbf e_x - \partial_y h \mbf e_y + \mbf e_z}{\sqrt{1+(\partial_x h)^2+(\partial_y h)^2}}.
\eeq

Second, the mean curvature $H$ encodes the deformation of the interface. It can be calculated by means of the height function 
$h(x,y,t)$ and is given by 
\beq
H=\dfrac{\left[ (1+h_y^2) h_{xx}-2 h_x h_y h_{xy}  +(1+ h_x^2)h_{yy} \right]}{2 (1+h_x^2+h_y^2)^{3/2}}\,,
\label{eq:curvature}
\eeq
where $h_x$ and $h_y$ denote partial derivatives of $h$ with respect to $x$ and $y$.

Third, the rate of strain tensor in the respective layer, $\hat{\mbf E}^{(i)}$, is given by
\beq
\hat{\mbf E}^{(i)}= \frac{1}{2}\left[ \nabla \tilde{ \mbf u}^{(i)} + \left(\nabla \tilde{\mbf u}^{(i)}\right)^T \right]\,, 
\label{eq:strain}
\eeq
where $\tilde{\mbf u}^{(i)}=\mbf u^{(i)} U_{vis}$.
The rate of strain tensor accounts for the normal viscous stresses at the interface. Without loss of generality, 
we restrict the discussion to the upper interface since velocities have a higher amplitude there and density 
stratification is less significant. Further calculations are done using dimensional quantities.

In the present framework, we  assumed a planar interface, i.e., $h=0$ results in $H=0$, which is appropriate
if the interface is pinned at the edges and if the right hand side of Eq.~\eqref{eq:normal:stress} remains small. 
Small should be understood in the sense that only deformations $\delta h$ of the interface are considered  
that are small compared to any length scale of the flow. A formal way to estimate the deformation 
is to perform  
a perturbation expansion of Eq.~\eqref{eq:normal:stress} with respect to $h$ as done in \citet{nepomnyashchy2006interfacial}
(see pp. 13 therein). This requires the calculation of the pressure from the zero-order solution of the Navier-Stokes equations.

Here, we will take a simpler approach by estimating a typical 
interface deformation as follows. Let us consider a fluid parcel
that moves downwards into the bulk of the lower layer away from the interface with velocity $U$. Simultaneously a column of 
denser liquid is present due to a temperature reduction by a magnitude of $\delta T$ in the top layer at this position. This causes 
a pressure difference which can be estimated by
\beq
\delta(\tilde p^{(2)} - \tilde p^{(3)}) \sim - \frac{1}{2}\rho^{(2)}U^2 - g\rho^{(3)}\beta_T^{(3)}\delta T  L_V\,,
\eeq 
where $L_V$ is the typical vertical variation scale. This process forces the interface to bulge downwards by $\delta h$ thus causing a negative 
curvature  $\delta H$. Simultaneously, the "hydrostatic contribution" to the pressure at the interface would rise by 
$g\delta h (\rho^{(2)}-\rho^{(3)})$. Now, one can relate these contributions via Eq.~\eqref{eq:normal:stress} and use the scaling relation 
for the viscous contribution, which is given by Eq.~\eqref{eq:strain}, and the curvature from Eq.\eqref{eq:curvature} to obtain 
\begin{align}
2\sigma^{II}_{ref} \frac{\delta h}{L_H^2}&\approx -\frac{1}{2}\rho^{(2)} U^2-g\rho^{(3)}\beta_T^{(3)} L_V \delta T\nonumber\\
                                                               &+g\delta h \left(\rho^{(3)}-\rho^{(2)}\right)- \mu^{(2)} \frac{U}{L_V}\,, 
\label{eq:defomation:1}
\end{align}
where the curvature term Eq.~\eqref{eq:curvature} is estimated with the help of a horizontal length scale $L_H$ and the viscous
contribution as well as the density contribution by means of a vertical length scale $L_V$. Finally, the interfacial deformation is 
estimated by rearranging \eqref{eq:defomation:1}. One obtains
\beq
\delta h = -\dfrac{ \dfrac{1}{2} \rho^{(2)} U^2 +\rho^{(3)}\beta_T^{(3)} L_V g \delta T + \mu^{(2)}  \dfrac{U}{L_V} }
{ \dfrac{2\sigma^{II}_{ref}}{L_H^2}+ g(\rho^{(2)}-\rho^{(3)})  }\label{eq:defomation:2}
\eeq
In order to estimate the deformation, we will adopt typical values of case H6. The velocity in the middle layer is $U\sim 3$mm/s 
(see Tab.~\ref{tab:results:dns}), the length scales $L_V\sim L_H\sim 10$mm as seen in Fig.~\ref{fig:var:height:2}. A typical temperature 
perturbation is $T^{II}\Theta \approx$ 1K. We note that the length unit for H6 is 40 mm. 
Inserting those values in Eq.~\eqref{eq:defomation:2} yields a $\delta h\approx$ -1.22 $\mu$m
which is considerably smaller than the smallest flow structures. In this view, the plane interface approximation is indeed appropriate 
and its errors should be insignificant compared to the other physical processes which have been neglected. For instance, vibrations 
acting on the system could trigger gravity waves.

Finally, we employ the estimate of the interface deformation by \citet{herreman2015tayler} (see their Eq. (5.7)) that equates the 
simulated kinetic energy with an increase in potential energy. This leads in our case to an estimate of
\beq
|\delta h|=\frac{ \frac{1}{2} \rho^{(2)} U^2 }{g(\rho^{(2)}-\rho^{(3)})}=0.66\mu {\rm m }.
\eeq
Thus we can conclude that no relevant dynamic interface deformation is triggered by thermal convection. The density difference 
between the layers is strong enough that any deformation is limited to a tiny magnitude. Furthermore, for an LMB application, 
the size of the middle layer  for H6 is comparatively large and would cause a high  voltage drop. Thus no higher velocities from the internal 
convection modes are to be expected. However, in systems with a tall upper layer and a strongly amplified RC mode, the deformation 
might become considerable again. In this case, we do not have reliable estimates of the velocity. Similar conclusions have been drawn 
by \citep{zikanov2015metal} for thermal convection and \citet{herreman2015tayler} for magnetohydrodynamic effects.

A curved equilibrium shape, which appears if the contact angle differs from 90 degrees, can have a particular influence on thermal 
convection. Such \textit{wall effects} should be studied together with the particular vessel geometry since the thermal properties 
of the battery vessel may induce horizontal temperature gradients, which in turn will induce Marangoni convection.

\section{Summary and Conclusions}
\label{sec:conclusion}
We have numerically studied thermal convection, induced by a uniform resistive heating, in a system of three liquid layers 
which serve as a simplified model for a liquid metal battery. Our model comprised the flow induced by interfacial tension 
gradients and gradients in density both of which are due to temperature gradients. The present LMB model 
has been simplified by disregarding electrochemical effects, the transport of mass across the separator layer, and the 
interaction with a vessel wall, respectively. Furthermore we exclude magnetohydrodynamic effects in connection with the
current density in the cell. Our main conclusions can be summarized as follows. 

(1) In the three-layer system Li||LiCl--KCl||Pb--Bi, driven by resistive heating, four main modes can drive convection. A 
fifth mode of anti-convection can appear as a secondary effect. The primary source for convection is the Rayleigh-B\'{e}nard 
convection mode due to internal heating in the molten salt layer. It obeys similar properties as a single convection layer subject 
to internal heating \cite{goluskin2016penetrative}. This process acts together with the second potential source of convective motion, 
the Marangoni convection effects at the upper interface. The typical thermal diffusivity ratios between a molten salt and a liquid metal 
lead to the appearance of Marangoni convection at the upper as well as at the lower interface of the separator to the electrodes
since heat is transported from the middle layer with significantly lower thermal diffusivity to the electrodes.
The fourth source for convection is  the unstable stratification in the upper electrode which causes a turbulent convection flow 
that is comparable to classical Rayleigh-B\'enard convection. Finally, anti-convection can increase the convective motion in the lower 
layer. For a fixed temperature drop, all modes will be damped when the thickness of the molten-salt layer is decreased, except the 
Rayleigh-B\'{e}nard convection mode in the top electrode.

(2) The exact linear stability thresholds for the onset of convection have been calculated for different system geometries and 
electrical current densities. By comparison with a full nonlinear DNS, it was found that those linear stability predictions indeed 
describe the conditions for the onset of convection well. The most critical mode in almost all of the simulated cases is the internal-
heating Rayleigh-B\'{e}nard mode. The stability threshold of this mode can be estimated by the middle layer Rayleigh number 
$\Ra^{(2)}$. Only for a particularly shallow middle layer, specifically for $d^{(2)}/d^{(1)}=0.1$ in our study, the classical Rayleigh-B\'{e}nard mode 
in the upper layer can become the main source for the onset of convection. 
The  threshold for this mode can be determined by the  Rayleigh number $\Ra^{(3)}$. 

(3) Several of our simulated configurations are also unstable with respect to Marangoni convection. Although this process is 
not the dominant mechanism of instability, it contributes significantly when both physical effects -- Rayleigh-B\'{e}nard and Marangoni 
effects -- are jointly enabled. The characteristic scale of the cellular flow patterns changes to a smaller length scale and the 
Nusselt number is increased simultaneously. The impact of the Marangoni effect on flow velocities is twofold. On the one hand, it is 
found that MC leads to a decrease of the velocity in the upper layer once the upper RC mode is active. On the other hand, for 
shallower middle layers, MC leads to a considerable increase of flow velocity in the lower and middle layers, respectively. To
conclude, although thermal MC turns out not to be the main source for convection, it affects the flow structures considerably.  
Further studies of convection in LMB should to our view include MC effects. 

\tk{We also expect that a number of changes will result from the presence of lateral walls, in particular when the aspect ratio of the 
cell (diameter/height) is small. Small aspect ratios will affect the cellular pattern formation and can thus suppress the heat 
transfer. This is known from RC \cite{bailon2010}. A finite thermal conductivity of the side walls will amplify the heat 
transfer, e.g. by a more pronounced thermal plume formation at the side walls \cite{verzicco2002}. Resulting horizontal temperature gradients can also 
drive buoyant as well as Marangoni convection. Finally, the wetting behavior of the liquid phase governs the interface geometry 
near the vessel walls. Additional corner modes could thus be amplified in connection with the heat transport through the walls.}

In future studies, the foundations of the physical model as well as the practical setup of prototypes will require significant advances.
The coupling between electrochemical reactions and the related ion transport has to be combined with the
thermal convection processes in order to receive a more precise and complete picture of the dynamical turbulent processes in 
LMBs.  Furthermore,  the assumption of a homogeneous current has to be improved since particular electrode geometries 
\citep{ning2015self} are employed in LMB prototypes where the coupling between to the reaction rates of the electrochemical 
reactions and convection processes can lead to heterogeneous heating. Moreover, the gradients in the composition of the 
lower electrode could lead to additional gradients in the surface tension since lithium is transferred across the interfaces. 
This might be an additional source for a stable density stratification. As we have discussed, some unknown material 
parameters, such as the interfacial tension, or resulting uncertainties limit the predictability of the importance of different convection 
processes. They ask for further experimental studies in the near future.

\begin{acknowledgments}
Financial support by the Deutsche Forschungsgemeinschaft in the framework of the Priority Programme SPP1506  is gratefully 
acknowledged. Furthermore, we thank the Computing Centre (UniRZ) of TU Ilmenau for access to its parallel computing resources. 
We also thank Oleg Zikanov and Norbert Weber for helpful discussions.
\end{acknowledgments}

\appendix

\section{Root mean squared velocity and Nusselt numbers}

In this appendix, we list the root mean square velocities and Nusselt numbers of the series H, JM and 
JS, respectively. They have been discussed in section IV. We exclude those runs which were linearly stable 
as well as the series JXS.
%-------------------------------------------------------------------------   
\begin{table*}
\begin{tabular}{ c c c c c c c c c c c}
\hline\hline
Case & \multicolumn{2}{c}{$u_{rms}^{(1)}$} & \multicolumn{2}{c}{$u_{rms}^{(2)}$} & \multicolumn{2}{c}{$u_{rms}^{(3)}$} & 
\multicolumn{2}{c}{$\Nu$} & $U_{vis}$ \\

&   $\;\;$RMC$\;\;$ & $\;\;$RC$\;\;$  & $\;\;$RMC$\;\;$ & $\;\;$RC$\;\;$& $\;\;$RMC$\;\;$ & $\;\;$RC$\;\;$    
          & $\;\;$RMC$\;\;$ & $\;\;$RC$\;\;$ & in $10^{-6}$ m/s \\
\hline          
H2 &\tk{4.26} & 3.80 & \tk{32.37}& 31.46    &  \tk{10.92} &9.97& \tk{1.49} & 1.45    &  12.9\\
H3 & \tk{15.32}& 15.26 & \tk{87.14}& 86.56    & \tk{31.20} &31.05 & \tk{2.22}  & 2.12	   &  8.6\\
H4 & \tk{31.40}& 37.14 & \tk{163.23} & 165.05 &\tk{56.13} &62.26 &\tk{2.88} &  2.67	   &  6.5 \\
H5 & \tk{38.86}& 35.81 & \tk{380.43} & 385.30 & \tk{226.43} &170.53 & \tk{4.45} &4.61 &  4.3\\
H6 & \tk{81.07}& 92.44&  \tk{704.04} & 709.52 & \tk{1180.94} &1359.30 & \tk{6.13} & 6.04 &  3.2 \\ \\

JM1 & \tk{2.60}& 2.06  & \tk{14.18} & 13.26  & \tk{4.81} &4.19  & \tk{1.07} &1.08  &  6.5 \\				
JM2 & \tk{7.44}& 6.67  & \tk{67.12} & 65.17  &  \tk{17.32} &15.84 & \tk{1.46} &1.42  &  6.5 \\	
JM3 &\tk{13.94} & 13.86 &\tk{111.51} & 108.37  & \tk{31.27} &30.25  & \tk{1.79} &1.69 &  6.5 \\				
JM4 &\tk{19.09} & 29.47 & \tk{208.71} & 210.72  &\tk{58.32} &141.46 & \tk{2.48} &2.15 &  6.5 \\ \\				
			
JS3 & \tk{6.88} & 6.30  & \tk{51.79}  & 48.57  &  \tk{13.19}  &11.56 & \tk{1.19}&1.16  &   6.5\\				
JS4 & \tk{11.72} & 8.78 & \tk{113.57} & 104.54 & \tk{26.20}  &20.48 & \tk{1.48}&1.39  &  6.5 \\
JS5 & \tk{43.63} & 26.71 &\tk{281.00} & 218.57 & \tk{478.55} &582.28& \tk{2.10}&1.79 & 6.5 \\				
\hline
\noalign{\smallskip}\hline
\end{tabular}
\caption{Statistical quantities which are averaged over the time interval $\tau_{av}$. We list root mean square velocities 
in all three layers as well as the Nusselt number. The values in column RMC and RC (Ma=0) belong to the DNS which are listed in Tab.~\ref{tab:paramteric-study:height}. The last column gives the velocity unit for a system with dimension listed in Tab.~\ref{tab:paramteric-study:height} } 
\label{tab:results:dns}
\end{table*}
%-------------------------------------------------------------------------   

\end{document}